\documentclass[a4paper,11pt]{article}
\usepackage{pos}

\newcommand{\ben}{\begin{eqnarray}}
\newcommand{\een}{\end{eqnarray}}
\newcommand{\la}{\langle}
\newcommand{\ra}{\rangle}

\title{Quantum Field Theories with Tensor Renormalization Group}

\ShortTitle{QFTs with TRG}

\author*[a]{Shinichiro Akiyama}
\author*[b]{Yoshinobu Kuramashi}
\author*[b]{Yusuke Yoshimura}

\affiliation[a]{Graduate School of Pure and Applied Sciences, University of Tsukuba, Tsukuba, Ibaraki
    305-8571, Japan}

\affiliation[b]{Center for Computational Sciences, University of Tsukuba, Tsukuba, Ibaraki
    305-8577, Japan}

\emailAdd{akiyama@het.ph.tsukuba.ac.jp}
\emailAdd{kuramasi@het.ph.tsukuba.ac.jp}
\emailAdd{yoshimur@ccs.tsukuba.ac.jp}

\notes{
\note{This is a combined contribution of ``Tensor renormalization group approach to (1+1)-dimensional Hubbard model'' by S.~Akiyama, ``Restoration of chiral symmetry in cold and dense Nambu$-$Jona-Lasinio model with tensor renormalization group'' by Y.~Kuramashi and ``Tensor renormalization group study of two-dimensional U(1) lattice gauge theory with a $\theta$ term'' by Y.~Yoshimura.}
\note{UTHEP-763, UTCCS-P-142}
}

\abstract{We report recent progress on the application of the tensor renormalization group (TRG) to quantum field theories pursued by the Tsukuba group. We explain how to treat the scalar, fermion, and gauge theories with the TRG method presenting the results for the phase transitions in the (3+1)-dimensional ((3+1)$d$) complex $\phi^4$ theory at finite density, (1+1)$d$ pure U(1) lattice gauge theory with a $\theta$ term, (3+1)$d$ Nambu--Jona-Lasinio model at finite density and (1+1)$d$ and (2+1)$d$ Hubbard models at an arbitrary chemical potential. It is demonstrated that the TRG method is free from the sign problem in practical calculations and applicable to the four-dimensional models.}

\FullConference{%
 The 38th International Symposium on Lattice Field Theory, LATTICE2021
  26th-30th July, 2021
  Zoom/Gather@Massachusetts Institute of Technology
}

\begin{document}

\maketitle

\section{Introduction}
In 2007 the tensor renormalization group (TRG) method was originally proposed to study two-dimensional (2$d$) classical spin systems in the field of condensed matter physics~\cite{Levin:2006jai}.\footnote{In this paper the TRG method or the TRG approach refers to not only the original numerical algorithm proposed by Levin and Nave \cite{Levin:2006jai} but also its extensions \cite{PhysRevB.86.045139,Shimizu:2014uva,Sakai:2017jwp,Adachi:2019paf,Kadoh:2019kqk,Akiyama:2020soe,adachi2020bondweighted,Kadoh:2021fri}.} This work attracted the attention of elementary particle physicists since the TRG method has several advantages over the Monte Carlo method. (i) The TRG method is  a deterministic numerical method so that it intrinsically does not have the sign problem encountered in stochastic methods including the standard Monte Carlo simulations.
(ii) The logarithmic dependence of the computational cost on the system size enables us to access the thermodynamic limit and the zero-temperature limit.
(iii) The TRG method allows direct manipulation of the Grassmann variables, which results in comparable computational costs between the fermionic and bosonic systems. It should be noted that we do not need to introduce the auxiliary fields to treat the four-fermi interactions, which are required in the Monte Carlo-based algorithms. 
(iv) We can obtain the partition function or the path integral itself. A typical benefit is the calculation of the pressure required in the equation of state, which is just given by the grand potential for the vast homogeneous system.

Unfortunately, there exist difficulties in the application of the TRG method to the quantum field theories (QFTs). For the scalar theories, we need to regularize the continuous degrees of freedom in the path-integral formalism. The gauge theories may have an additional difficulty to treat the redundant degrees of freedom due to the local gauge symmetry. The fermion fields are expressed with the Grassmann variables in the path-integral formalism so that we need to incorporate the Grassmann algebra in the TRG method. Furthermore, we need an efficient  algorithm to calculate higher-dimensional theories, since the original TRG algorithm~\cite{Levin:2006jai} is applicable to only two-dimensional (2$d$) models.
In this report, we explain how we have overcome these difficulties and present some physics results that the current Monte Carlo methods would never achieve due to the sign problem or the computational cost.

This report consists of two parts. We first discuss the application of the TRG method to the bosonic systems. We give a brief review of the analysis of the (3+1)$d$ complex $\phi^4$ theory at finite density with the TRG method in Sec.~\ref{subsec:scalar}. It is instructive to demonstrate how to treat the continuous degree of freedom in the scalar theories and show evidence that the TRG method is free from the sign problem. In Sec.~\ref{subsec:u1gauge}, we present the results for the 2$d$ U(1) gauge theory with a $\theta$ term, which is another notorious example with the complex action problem. The second part is devoted to discussing the fermionic systems. In Sec.~\ref{subsec:gtrg}, we briefly explain how to apply the TRG method to evaluate fermionic path integrals. Section~\ref{subsec:njl} presents the TRG study of the Nambu$-$Jona-Lasinio (NJL) model in the cold and dense region as a representative case of the fermionic systems. Based on a similarity between the NJL model and the Hubbard model, we also show the applicability of the TRG method to the (1+1)$d$ and (2+1)$d$ Hubbard models in Secs.~\ref{subsec:hubbard_1+1} and \ref{subsec:hubbard_2+1}.  Summary and outlook are given in Sec.~\ref{sec:summary}.

\section{Bosonic systems}

\subsection{(3+1)$d$ complex $\phi^4$ theory at finite density}
\label{subsec:scalar}

The QFT application of the TRG method was first tried to the (1+1)$d$ real scalar $\phi^4$ theory in 2012, where the spontaneous $\mathbb{Z}_2$ symmetry breaking was investigated by employing an expansion method with the orthogonal functions to regularize the continuous degrees of freedom for the scalar field~\cite{Shimizu:2012zza}. Several years later this model was revisited employing the Gauss quadrature to make a different regularization of the continuous scalar fields and succeeded in determining the critical coupling in the continuum limit~\cite{Kadoh:2018tis}. This work was followed by the study of the (1+1)$d$ complex $\phi^4$ theory at finite density, which is a typical system with the complex action problem. 
The Silver Blaze phenomenon, where bulk observables are independent of the chemical potential $\mu$ up to some critical point $\mu_{\rm c}$ in the thermodynamic limit at zero temperature, was successfully confirmed on the extremely large volume of $1024^2$ demonstrating that the TRG method does not suffer from the complex action problem~\cite{Kadoh:2019ube}. In this subsection, we present the recent results for the (3+1)$d$ complex $\phi^4$ theory at finite density~\cite{Akiyama:2020ntf} explaining how to regularize the continuous scalar fields with the Gauss quadrature.

The (3+1)$d$ complex $\phi^4$ theory at finite density, which is defined by a complex action, is expected to show the Silver Blaze phenomenon. Since the complex phase of the action plays an essential role in this phenomenon, this model has been studied by various methods intended to overcome or tame the sign problem, such as the complex Langevin approach~\cite{Aarts:2008wh}, the thimble method~\cite{Cristoforetti:2013wha,Fujii:2013sra,Mori:2017nwj}, and the world-line representation~\cite{Gattringer:2012df,Orasch:2017niz}.
We explain how to define a finite-dimensional tensor with regularization of scalar fields and show that the efficiency of the TRG method to investigate the Silver Blaze phenomena without suffering from the sign problem.

\subsubsection{Tensor network representation with the Gauss quadrature}

The lattice action of the (3+1)$d$ complex $\phi^4$ theory at finite density is defined by
\begin{align}
	\label{eq:action_phi4}
	S[\phi]=\sum_{n\in\Lambda}	\left[ (8+m^2)|\phi_n|^2 +\lambda|\phi_n|^4
		-\sum_{\nu=1}^4 \left( {\rm e}^{\mu\delta_{\nu 4}}\phi_n^\ast\phi_{n+\hat\nu}
			+{\rm e}^{-\mu\delta_{\nu 4}}\phi_n\phi_{n+\hat\nu}^\ast \right) \right]
\end{align}
with the complex scalar field $\phi_n$, the bare mass $m$, the coupling constant $\lambda>0$ and the chemical potential $\mu$.
$\phi_n$ lives on a site $n=(n_1,n_2,n_3,n_4)\in\Lambda(\subset\mathbb{Z}^4)$.
The lattice spacing has been set to 1.
We choose the periodic boundary condition for the scalar field: $\phi_{n+N_{\nu}{\hat \nu}}=\phi_n$ for $\nu=1,2,3,4$ with ${\hat \nu}$ is the unit vector of the $\nu$-direction.

Let us derive the tensor network representation of the path integral,
\begin{align}
	Z=\int\mathcal D\phi \, {\rm e}^{-S[\phi]}.
\end{align}
We employ the polar coordinate $\phi_n(r_n,\theta_n)=r_n {\rm e}^{{\rm i}\pi\theta_n}$ and the associated integral measure is given by
\begin{align}
	\int\mathcal D\phi =
	\prod_{n\in\Lambda} \int_0^\infty {\rm d}r_n r_n \int_{-1}^1 \pi {\rm d}\theta_n.
\end{align}
In general, an integral of a function $f(\varphi)$ can be evaluated via the Gauss quadrature rule, 
\begin{align}
	\int {\rm d}\varphi f(\varphi)
	\approx \sum_{\alpha=1}^K w_\alpha f\left( \varphi^{(\alpha)} \right),
	\label{eq:GLQ}
\end{align}
where $\varphi^{(\alpha)}$ and $w_\alpha$ are the $\alpha$-th node of the $K$-th polynomial and the associated weight, respectively.
Now, the continuous variables $r_n$ and $\theta_n$ are regularized by the $K_1$-point Gauss-Laguerre and $K_2$-point Gauss-Legendre quadrature rule, respectively.
$r_{\alpha}$ and $w_{1,\alpha}$ denote the $\alpha$-th node and weight in the former quadrature and 
$\theta_{\beta}$ and $w_{2,\beta}$ are for the $\beta$-th node and its weight in the latter one.
The regularized path integral is given by
\begin{align}
	Z(K_1,K_2)= \sum_{ \{\alpha,\beta\} } \left[ \prod_{n\in\Lambda} 
		(w_{1,\alpha_n} {\rm e}^{r_{\alpha_n}} r_{\alpha_n}) ( \pi w_{2,\beta_n} ) \right]
	{\rm e}^{-S[\phi(r_\alpha,\theta_\beta)]}
\end{align}
with
\begin{align}
	\sum_{ \{\alpha,\beta\} }=
	\prod_{n\in\Lambda} \sum_{\alpha_n=1}^{K_1} \sum_{\beta_n=1}^{K_2}.
\end{align}
Introducing the $(K_1 K_2)\times (K_1 K_2)$  square matrices, 
\begin{multline}
		M^{[\nu]}_{\alpha\beta,\alpha'\beta'}
		=\sqrt[4]{\pi}
		\sqrt[8]{r_\alpha w_{1,\alpha} w_{2,\beta} r_{\alpha'} w_{1,\alpha'} w_{2,\beta'}}
		\exp\left( \frac{r_\alpha+r_{\alpha'}}{8} \right)
		\\
		\cdot \exp\left[ \left( 1+\frac{m^2}{8} \right)\left( r_\alpha^2+r_{\alpha'}^2 \right)
		+\frac{\lambda}{8} \left( r_\alpha^4 +r_{\alpha'}^4 \right)
		-2 r_\alpha r_\beta \cos(\pi(\theta_\beta-\theta_{\beta'})-{\rm i}\mu\delta_{\nu 4}) \right],
\end{multline}
the approximated path integral $Z(K_1,K_2)$ is expressed as
\begin{align}
	Z(K_1,K_2)
	=\sum_{ \{\alpha,\beta\} } \prod_{n\in\Lambda} \prod_{\nu=1}^4
	M^{[\nu]}_{\alpha_n\beta_n,\alpha_{n+\hat\nu}\beta_{n+\hat\nu}}.
\end{align}
We then apply the singular value decomposition (SVD) to each matrix $M$: 
\begin{align}
	M^{[\nu]}_{\alpha\beta,\alpha'\beta'}
	=\sum_{k=1}^{K_1 K_2}
	U^{[\nu]}_{\alpha\beta,k} \sigma_k^{[\nu]} V^{[\nu]*}_{\alpha'\beta',k}
	\approx \sum_{k=1}^D
	U^{[\nu]}_{\alpha\beta,k} \sigma_k^{[\nu]} V^{[\nu]*}_{\alpha'\beta',k},
\label{eq:m_svd}
\end{align}
where $\sigma_k^{[\nu]}$ is the $k$-th singular value sorted in the descending order,
and $U^{[\nu]}$ and $V^{[\nu]}$ are the unitary matrices composed of the singular vectors. The truncation parameter $D(<K_{1}K_{2})$ is chosen as the bond dimension in the TRG algorithm.
Finally, the path integral is approximately represented by the tensor network as
\begin{align}
	\label{eq:Z_phi4}
	Z(K_1,K_2)
	=\sum_{x,y,z,t} \prod_{n\in\Lambda}
	T_{x_n y_n z_n t_n x_{n-\hat 1} y_{n-\hat 2} z_{n-\hat 3} t_{n-\hat 4} },
\end{align}
where the tensor $T$ is defined by
\begin{align}
	T_{i_1 i_2 i_3 i_4 j_1 j_2 j_3 j_4 }
	=\sum_{\alpha=1}^{K_1} \sum_{\beta=1}^{K_2} \prod_{\nu=1}^4
	\sqrt{\sigma^{[\nu]}_{i_\nu}\sigma^{[\nu]}_{j_\nu}}
	U^{[\nu]}_{\alpha\beta,i_\nu} V^{[\nu]*}_{\alpha\beta,j_\nu}.
\end{align}

\subsubsection{Numerical setup}

We choose $m=0.1$ and $\lambda=1.0$ for the lattice complex $\phi^4$ theory of Eq.~\eqref{eq:action_phi4}. 
The path integral of Eq.~\eqref{eq:Z_phi4} is evaluated using the anisotropic TRG (ATRG) algorithm \cite{Adachi:2019paf} on a periodic lattice with the volume $V=L^4$ ($L=2^m, m \in \mathbb{N}$). The bond dimension is set to $D=45$ and the polynomial orders in the Gauss quadrature methods to  $K=K_{1}=K_{2}=64$. Convergence with respect to these algorithmic parameters is checked in Ref.~\cite{Akiyama:2020ntf}.

\subsubsection{Silver Blaze phenomenon}

\begin{figure}[htbp]
	\begin{minipage}[t]{0.48\hsize}
    		\centering
   	 	\includegraphics[width=1.0\hsize]{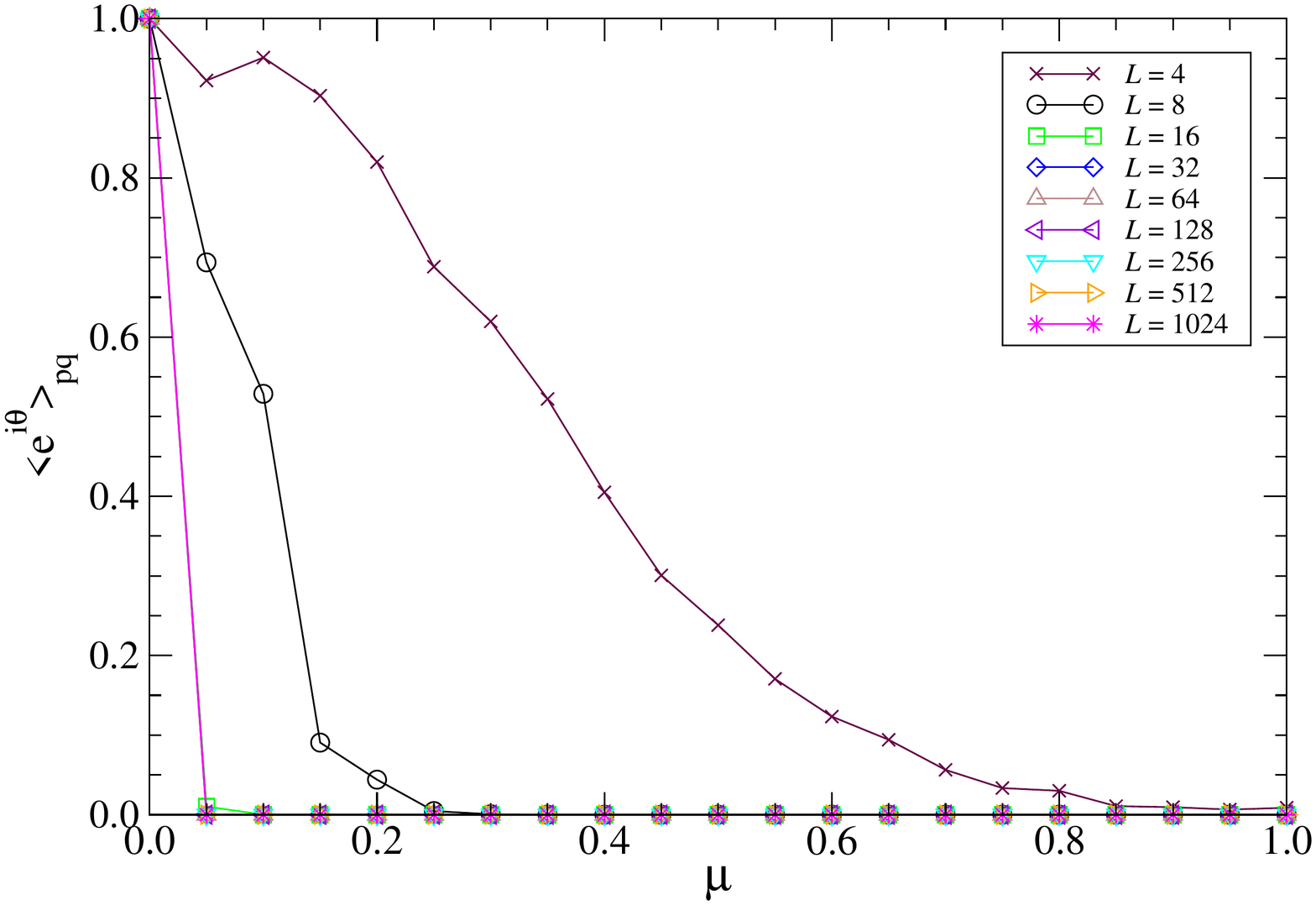}
  		\caption{Average phase factor as a function of $\mu$ with $m^2 = 0.01$, $\lambda=1.0$, $K=64$, $D=45$. The lattice volume $V$ is varied from $4^4$ to $1024^4$.}
  		\label{fig:average_phase}
  	\end{minipage}
  	\hspace*{3mm}
	\begin{minipage}[t]{0.48\hsize}
    		\centering
    		\includegraphics[width=1.0\hsize]{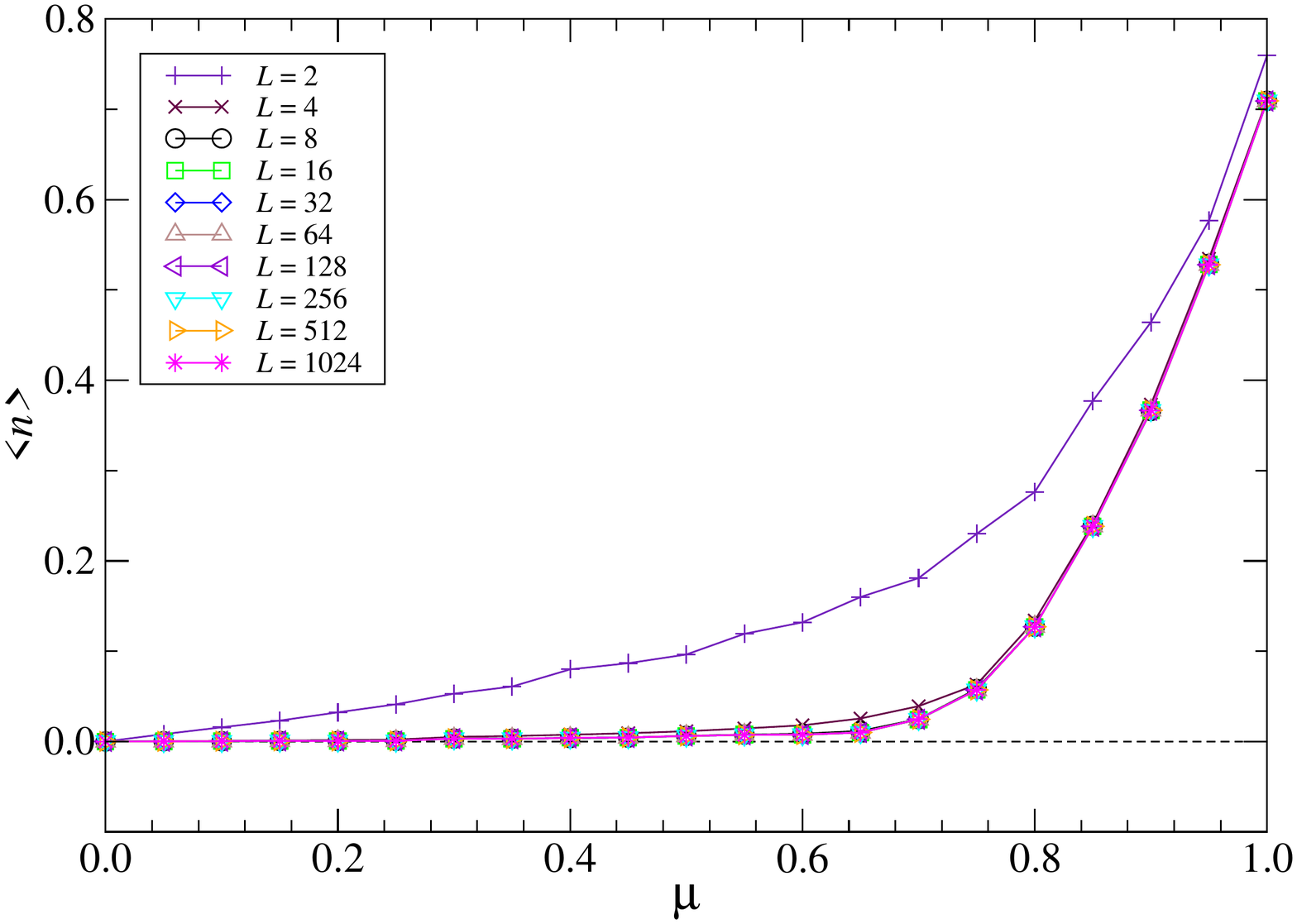}
  		\caption{Particle number density as a function of $\mu$ with the lattice volume varied from $2^4$ to $1024^4$. Other parameters of $m$, $\lambda$, $K$ and $D$ are the same as those in Fig.~\ref{fig:average_phase}.}
  		\label{fig:cphi4_number_density}
	\end{minipage}
\end{figure}

We first define the phase-quenched path integral as
\begin{align}
	\label{eq:zpq}
  	Z_{\mathrm{pq}} = \int \mathcal{D} \phi \, {\rm e}^{-\mathrm{Re}\left( S \right)}, 
\end{align}
where only the real part of the Boltzmann factor is taken by the decomposition 
${\rm e}^{-S}={\rm e}^{-\mathrm{Re}\left( S \right)}{\rm e}^{{\rm i}\theta}$.
The expectation value of an observable ${\mathcal O}$ with the phase-quenched theory is expressed as  $ \langle {\mathcal O} \rangle_{\rm pq}$, which is related to $ \langle {\mathcal O} \rangle$ with the full theory as 
\begin{align}
	\label{ratio_full}
  	\langle {\mathcal O}\rangle
  	=
  	\frac{\langle {\mathcal O}{\rm e}^{{\rm i}\theta}\rangle_{\rm pq}}{\langle {\rm e}^{{\rm i}\theta}\rangle_{\rm pq}}.
\end{align}
In case that the phase factor oscillates frequently in the large $\mu$ region, it is difficult for the Monte Carlo method to evaluate the ratio because of the vanishing contributions from both the numerator and the denominator (This is the so-called sign problem). 
In Fig.~\ref{fig:average_phase} we plot the average phase factor $\langle {\rm e}^{{\rm i}\theta}\rangle_{\rm pq}={Z}{/Z_{\mathrm{pq}}}$ as a function of $\mu$ varying the lattice volume $V$. This quantity measures how severe the sign problem is for given parameters of $\mu$ and $V$. We observe that $\langle {\rm e}^{{\rm i}\theta}\rangle_{\rm pq}$ becomes close to zero as either of the volume or the chemical potential increases. On the largest volume of $V=1024^4$, which is essentially regarded as the thermodynamic limit at zero temperature, the average phase factor quickly falls off from one at $\mu=0$ to zero for $\mu\ge 0.05$, where the Monte Carlo method does not work.
In Fig.~\ref{fig:cphi4_number_density} we plot the $\mu$ dependence of 
the particle number density defined by
\begin{align}
	\langle n \rangle=\frac{1}{V}
  	\frac{\partial \ln Z}{\partial\mu},
\end{align}
which is evaluated by the ATRG algorithm with impurity tensors \cite{Kadoh:2018tis}.  We observe that the Silver Blaze phenomenon becomes manifest on the larger volume toward the thermodynamic limit at zero temperature: the particle number density stays around zero up to $\mu_{\rm c}\approx 0.65$ and shows rapid increase beyond $\mu_{\rm c}$, even in the regime with the vanishing $\langle {\rm e}^{{\rm i}\theta}\rangle_{\rm pq}$.

\subsection{(1+1)$d$ pure U(1) lattice gauge theory with a $\theta$ term}
\label{subsec:u1gauge}

In comparison with the scalar theories, it is more difficult to develop an efficient TRG algorithm for gauge theories because of the redundancy of gauge degrees of freedom. 
So far a few numerical attempts have been made to investigate the phase transition in the pure lattice gauge theories \cite{Kuramashi:2019cgs,Kuramashi:2018mmi}. 
Here we propose to use the Gauss quadrature to regularize the continuous gauge theories~\cite{Kuramashi:2019cgs}. This is motivated by the future application of the TRG method to the $(3+1)d$ SU($N$) gauge theories. 

The (1+1)$d$ pure U(1) lattice gauge theory with a $\theta$ term is the simplest pure lattice gauge theory with a $\theta$ term. There are two motivations to study it with the TRG method. Firstly, this model is a case of the complex action due to the $\theta$ term. The analytical result for the partition function is already known~\cite{Wiese:1988qz}: This model undergoes the first-order phase transition at $\theta=\pi$. 
It is worth noting that a recent numerical study with the complex Langevin approach finds that the naive implementation fails for this theory~\cite{Hirasawa:2020bnl}.
Therefore, it should be a good testbed to check that the TRG method does not suffer from the complex action problem or the sign problem. 
Secondly, we try to apply the Gauss quadrature method with some improvement to discretize the phase in the U(1) link variable. This follows the success of the Gauss quadrature method to discretize the continuous degree of freedom in the scalar theories~\cite{Kadoh:2018tis,Kadoh:2019ube}.

\subsubsection{Tensor network representation with the Gauss quadrature}

The Euclidean action of the (1+1)$d$ pure U(1) lattice gauge theory with a $\theta$ term is defined by
\begin{gather}
	S= -\beta \sum_{x\in\Lambda} \cos p_x -{\rm i}\theta Q, \\
	p_x= \varphi_{x,1}+\varphi_{x+\hat 1,2}-\varphi_{x+\hat 2,1}-\varphi_{x,2}, \\
	Q= \frac{1}{2\pi} \sum_{x\in\Lambda} q_x, \quad q_x=p_x \bmod 2\pi,
\end{gather}
where $\varphi_{x,\nu}\in [-\pi,\pi]$ is the phase of U(1) link variable at site $x$ in $\nu$ direction.
The range of $q_x$ is $[-\pi,\pi]$ and it can be expressed as follows by introducing an integer $n_x$:
\begin{align}
	q_x =p_x +2\pi n_x, \quad n_x\in \{-2,-1,0,1,2\}.
\end{align}
For the periodic boundary condition, the topological charge $Q$ becomes an integer:
\begin{align}
	Q =\sum_{x\in\Lambda} \left( \frac{p_x}{2\pi} +n_x \right) =\sum_{x\in\Lambda} n_x.
\end{align}
The tensor may be given with continuous indices,
\begin{align}
	\mathcal T(\varphi_{x,1},\varphi_{x+\hat 1,2},\varphi_{x+\hat 2,1},\varphi_{x,2}) 
	=\exp\left( \beta\cos p_x +{\rm i}\frac{\theta}{2\pi}q_x \right).
\end{align}
The partition function is represented as
\begin{align}
	Z =\left( \prod_{x\in\Lambda}\prod_{\nu=1,2}\int_{-\pi}^\pi \frac{{\rm d}\varphi_{x,\mu}}{2\pi} \right) 
	\prod_{x\in\Lambda} \mathcal T(\varphi_{x,1},\varphi_{x+\hat 1,2},\varphi_{x+\hat 2,1},\varphi_{x,2}).
	\label{eq:PF}
\end{align}
We regularize all the integrals in Eq.~\eqref{eq:PF} using the Gauss-Legendre quadrature with the polynomial order $K$. The finite-dimensional tensor network is expressed as
\begin{align}
	Z(K) \approx \sum_{\{\alpha\}}\prod_{x\in\Lambda} T_{\alpha_{x,1}\alpha_{x+\hat 1,2}\alpha_{x+\hat 2}\alpha_{x,2}}
\end{align}
with the discretized local tensor 
\begin{align}
	T_{ijkl} =\frac{\sqrt{w_i w_j w_k w_l}}{(2\pi)^2}
	\mathcal T\left( \varphi^{(i)},\varphi^{(j)},\varphi^{(k)},\varphi^{(l)} \right).
\end{align}

\subsubsection{Improvement technique to reduce the truncation error}

We have developed further improvements for the above method.
In the SVD procedure to prepare the initial tensor before starting the iterative TRG steps \cite{Kadoh:2018hqq,Kadoh:2018tis,Kadoh:2019ube},
we employ the following eigenvalue decomposition:
\begin{align}
	M_{ijkl}=\frac{\sqrt{w_i w_j w_k w_l}}{(2\pi)^4}\int_{-\pi}^\pi {\rm d}\varphi_1 {\rm d}\varphi_2
	\mathcal T \left( \varphi^{(i)},\varphi^{(j)},\varphi_1,\varphi_2 \right)
	\mathcal T^\ast \left( \varphi^{(k)},\varphi^{(l)},\varphi_1,\varphi_2 \right),
	\label{eq:EVD}
\end{align}
which is essentially equivalent to 
\begin{align}
	M_{ijkl} =\lim_{K^\prime\rightarrow\infty} \sum_{m,n=1}^{K^\prime} T_{ijmn}T^\ast_{klmn}.
\end{align}
This procedure is expected to reduce the discretization errors in $M_{ijkl}$.
To evaluate Eq.~\eqref{eq:EVD}, we use the character expansion~\cite{Imachi:1997fy,Hassan:1995dn}:
\begin{align}
	\mathcal T(\varphi_1,\varphi_2,\varphi_3,\varphi_4) 
	=\sum_{m,n=-\infty}^\infty {\rm e}^{{\rm i}n(\varphi_1+\varphi_2-\varphi_3-\varphi_4)}
	I_m(\beta) J_{n-m}(\theta)
\end{align}
where $I_m(\beta)$ is the $m$-th order modified Bessel function of the first kind and
\begin{align}
	J_n(\theta) =(-1)^n \frac{2}{\theta+2\pi n} \sin\left(\frac{\theta}{2}\right).
\end{align}
Then, Eq.~\eqref{eq:EVD} is rewritten as
\begin{align}
	M_{ijkl}
	=\frac{\sqrt{w_i w_j w_k w_l}}{(2\pi)^4}
	\sum_{n=-\infty}^\infty {\rm e}^{{\rm i}n(\varphi^{(i)}+\varphi^{(j)}-\varphi^{(k)}-\varphi^{(l)})}
	\left( \sum_{m,m^\prime=-\infty}^\infty I_m(\beta)I_{m^\prime}(\beta) J_{n-m}(\theta)J_{n-m^\prime}(\theta) \right).
\end{align}
In the practical calculation,
the sums of $n,m$ and $m^\prime$ can be truncated when the contributions of the terms are small enough. In this work we discard the contributions of $I_{m,m^\prime}/I_0 < 10^{-12}$ or $J_{n-m,n-m^\prime}/J_0 < 10^{-12}$.

\subsubsection{Numerical setup}

The partition function of Eq.~\eqref{eq:PF} is evaluated with the TRG algorithm at $\beta=$0.0 and 10.0 as a function of $\theta$ on a $V=L\times L$ lattice, where $L$ is enlarged up to 1024.
We choose $K=32$ for the polynomial order of the Gauss-Legendre quadrature.
The SVD procedure in the TRG algorithm is truncated with $D=32$.
We have checked that these choices of $D$ and $K$ provide us sufficiently converged results for all the parameter sets employed in this work.
Since the scaling factor of the TRG method is $\sqrt{2}$, allowed lattice sizes for the partition function are $L=\sqrt{2},2,2\sqrt{2},\cdots,512\sqrt{2},1024$.
The periodic boundary condition is employed in both directions so that the topological charge $Q$ is quantized to be an integer. 

\subsubsection{Free energy and topological charge density}

\begin{figure}[htbp]
  \begin{minipage}[t]{0.48\hsize}
    \centering
    \includegraphics[width=1.0\hsize]{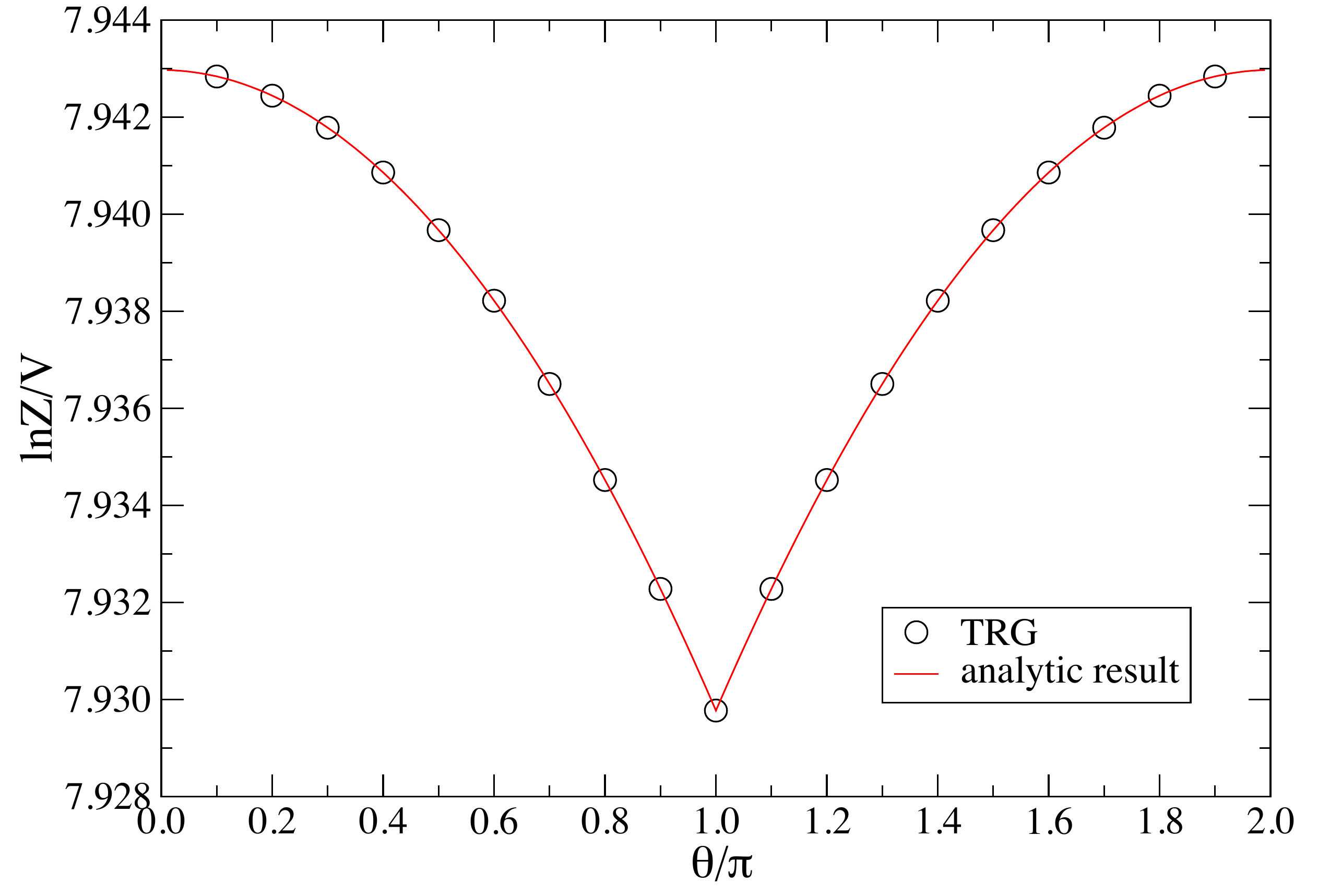}
    \caption{$\theta$ dependence of free energy  at $\beta=10.0$ with $K=32$ and $D=32$ on a $1024\times 1024$ lattice. Solid curve denotes the analytic result of Eq.~(\ref{eq:z_anal_u1theta})}
    \label{fig:freeenergy}
  \end{minipage}
  \hspace*{3mm}
\begin{minipage}[t]{0.48\hsize}
    \centering
    \includegraphics[width=1.0\hsize]{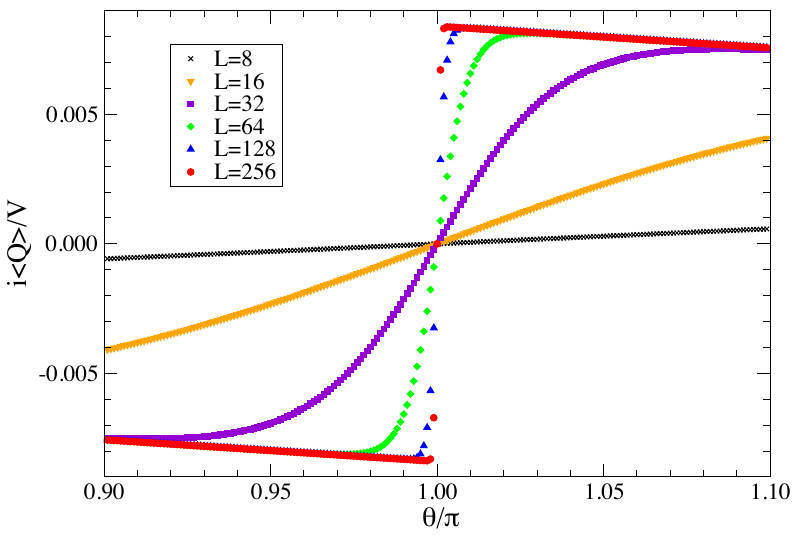}
    \caption{Topological charge density with $8\le L\le 256$ as a function of $\theta$ at $\beta=10.0$.}
    \label{fig:tc}
\end{minipage}
\end{figure}

The analytic result for the partition function of Eq.~\eqref{eq:PF} is given by \cite{Wiese:1988qz}:
\ben
&&Z_{\rm analytic}=\sum_{Q=-\infty}^{\infty} \left(z_{\rm P}(\theta+2\pi Q,\beta)\right)^V,
\label{eq:z_anal_u1theta}\\
&&z_{\rm P}(\theta,\beta)=\int_{-\pi}^{\pi}\frac{{\rm d} \varphi_{\rm P}}{2\pi}\exp\left(\beta\cos \varphi_{\rm P} +{\rm i}\frac{\theta}{2\pi}\varphi_{\rm P}\right),
\een
where $z_{\rm P}(\theta,\beta)$ denotes the one-plaquette partition function with $\varphi_{\rm P}\in [-\pi,\pi]$.
In Fig.~\ref{fig:freeenergy} we compare our numerical results for the free energy $\ln Z/V$ with the above exact results as a function of $\theta$. We observe a good consistency over the range of $0\le \theta \le 2\pi$. The kink of the free-energy at $\theta=\pi$ indicates the first-order phase transition.

The expectation value of the topological charge $\la Q\ra$ at $\beta=10.0$ is obtained by the numerical derivative of the free energy with respect to $\theta$:
\ben
&&\la Q\ra=-{\rm i}\frac{\partial \ln Z}{\partial \theta}.
\een 
Figure~\ref{fig:tc} shows the volume dependence of $\la Q\ra/V$ around $\theta=\pi$ with much finer resolution of $\theta$ than Fig.~\ref{fig:freeenergy}, where the first-order phase transition is expected. We observe that a finite discontinuity emerges with mutual crossings of curves between different volumes at $\theta=\pi$ as the lattice size $V=L^2$ is increased. This feature indicates this system undergoes a first-order phase transition  at $\theta=\pi$.

\subsubsection{Topological susceptibility}

\begin{figure}[htbp]
  \begin{minipage}[t]{0.48\hsize}
    \centering
    \includegraphics[width=1.0\hsize]{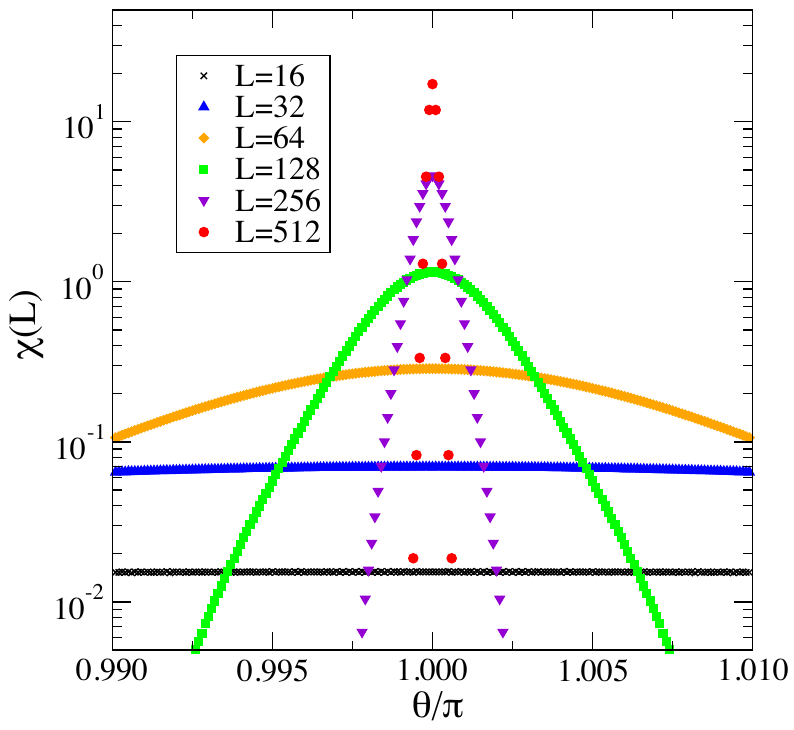}
    \caption{Topological susceptibility $\chi(L)$ as a function of $\theta$ with $16\le L\le 512$.}
    \label{fig:chi}
  \end{minipage}
  \hspace*{3mm}
\begin{minipage}[t]{0.48\hsize}
    \centering
    \includegraphics[width=1.0\hsize]{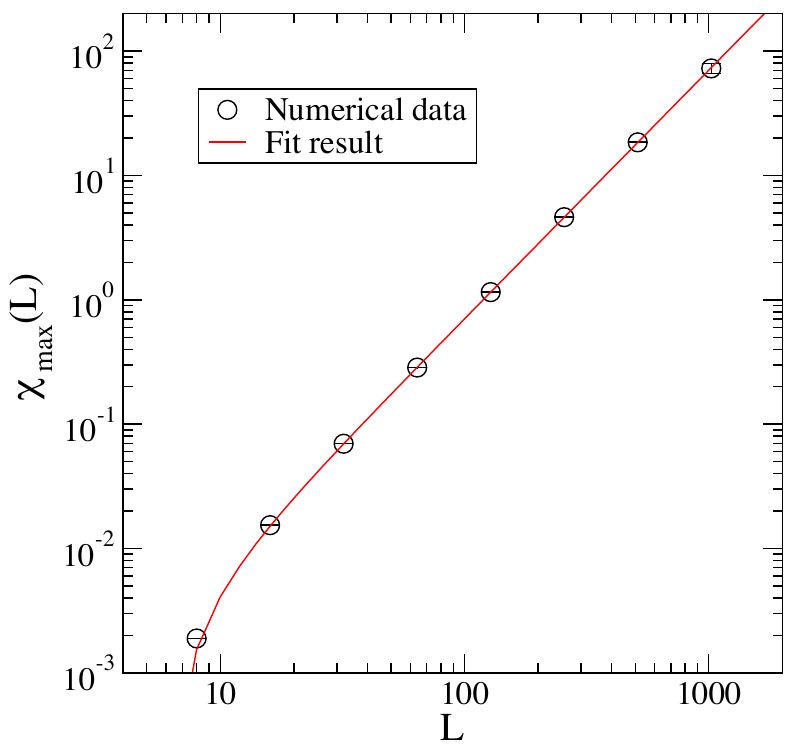}
    \caption{Peak height of topological susceptibility $\chi_{\rm max}(L)$ as a function of $L$. Solid curve denotes the fit result.}
    \label{fig:chi_max_L}
\end{minipage}
\end{figure}

We investigate the properties of the phase transition by applying the finite size scaling analysis to the topological susceptibility: 
\ben
&&\chi(L)=-\frac{1}{V} \frac{\partial^2 \ln Z}{\partial \theta^2}.
\label{eq:chi}
\een 
Figure~\ref{fig:chi} shows the topological susceptibility  as a function of $\theta$ for various lattice sizes. The
peak structure is observed and its height $\chi_{\rm max}(L)$ grows as $L$ increases.
In order to determine the peak position $\theta_{\rm c}(L)$ and the peak height $\chi_{\rm max}(L)$ at each $L$, we employ the quadratic approximation of the 
topological susceptibility around the peak position:
\ben
&&\chi(L)\sim \chi_{\rm max}(L)+R\left(\theta-\theta_{\rm c}(L)\right)^2
\label{eq:chi_fit}
\een
with $R$ a constant.
We expect that the peak height scales with $L$ as
\ben
\chi_{\rm max}(L) \propto L^{\gamma/\nu},
\een
where $\gamma$ and $\nu$ are the critical exponents. The $L$ dependence of the peak height $\chi_{\rm max}(L)$ is plotted in Fig.~\ref{fig:chi_max_L}. The solid curve represents the fit result obtained with the fit function of $\chi_{\rm max}(L) = A+ B L^{\gamma/\nu}$ choosing the fit range of $128\le L\le 1024$. The results for the fit parameters are given by $A=-3(2)\times 10^{-3},B=7.12(8)\times 10^{-5}$ and $\gamma/\nu=1.998(2)$. The value of the exponent $\gamma/\nu$ is consistent with two, which is the expected critical exponent in the first-order phase transition in the two-dimensional system.

\section{Fermionic systems}
\label{sec:fermion}

In 2014, the TRG method was applied to analyze the phase structures of the lattice Schwinger model with and without the $\theta$ term, which contains the sign problem, developing the Grassmann TRG (GTRG) algorithm~\cite{Shimizu:2014uva,Shimizu:2014fsa}. This was the first numerical calculation of the fermionic system with the TRG method and also the first one for the gauge theory. The GTRG algorithm was also applied to the analysis of the finite-density (1+1)$d$ lattice Gross-Neveu model~\cite{Takeda:2014vwa}. After that, the Grassmann HOTRG (GHOTRG) algorithm was developed based on the GTRG algorithm in order to investigate the higher-dimensional fermionic systems in particle physics~\cite{Sakai:2017jwp}. The validity of the GHOTRG algorithm was tested using the Green functions with the (2+1)$d$ relativistic free Wilson fermion, whose exact values are analytically calculable~\cite{Yoshimura:2017jpk}. 

\begin{figure}[htbp]
	\begin{minipage}[t]{0.48\hsize}
    		\centering
   		 \includegraphics[width=1.0\hsize]{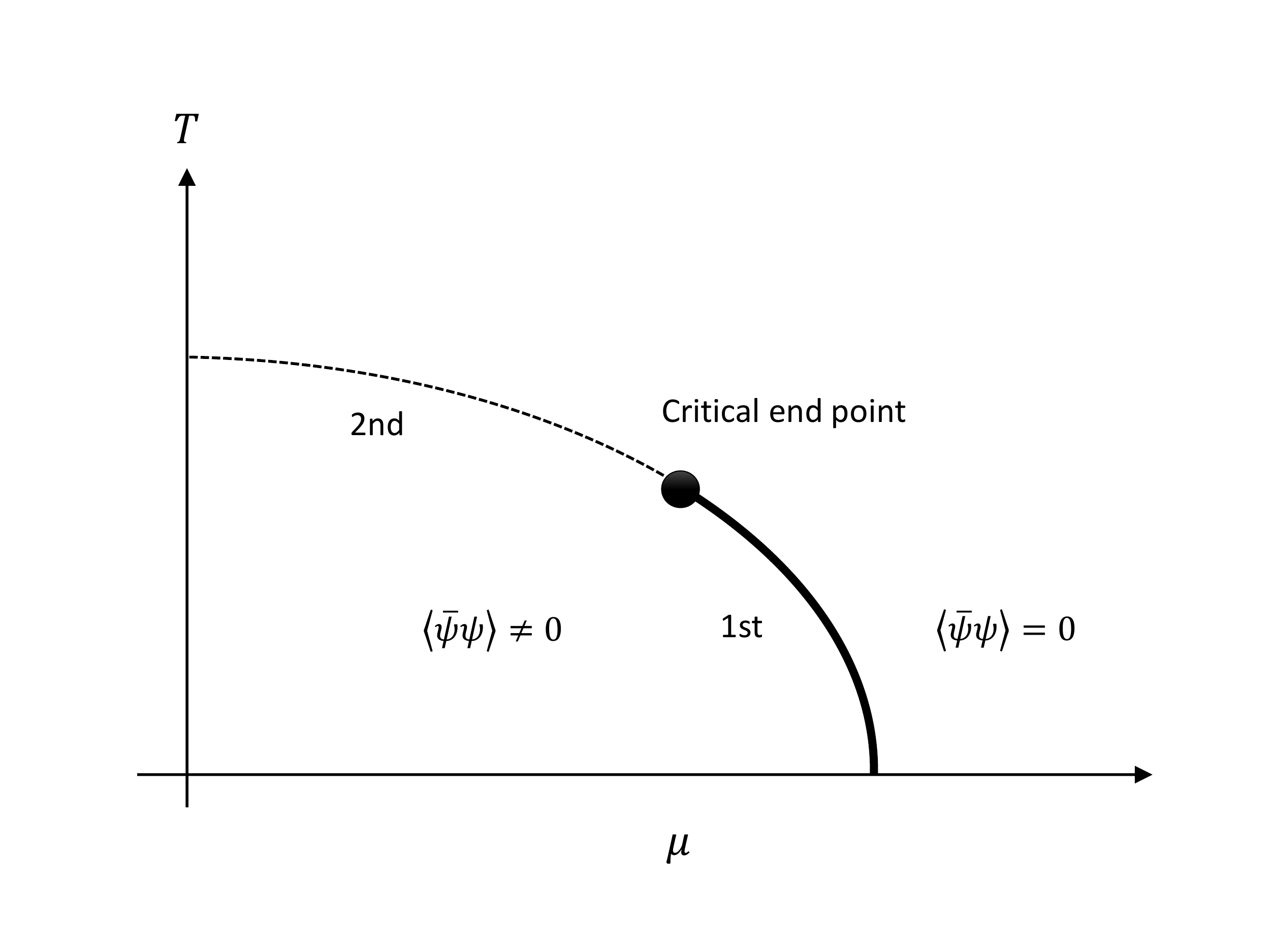}
    		\caption{Schematic view of expected phase diagram of the NJL model on the $T$-$\mu$ plane. Solid and broken curves represent the first- and second-order phase transitions, respectively. Closed circle denotes the critical end point (CEP) where the first-order phase transition line terminates.}
    		\label{fig:phdgm_njl}
 	 \end{minipage}
 	 \hspace*{3mm}
	\begin{minipage}[t]{0.48\hsize}
   		\centering
    		\includegraphics[width=1.0\hsize]{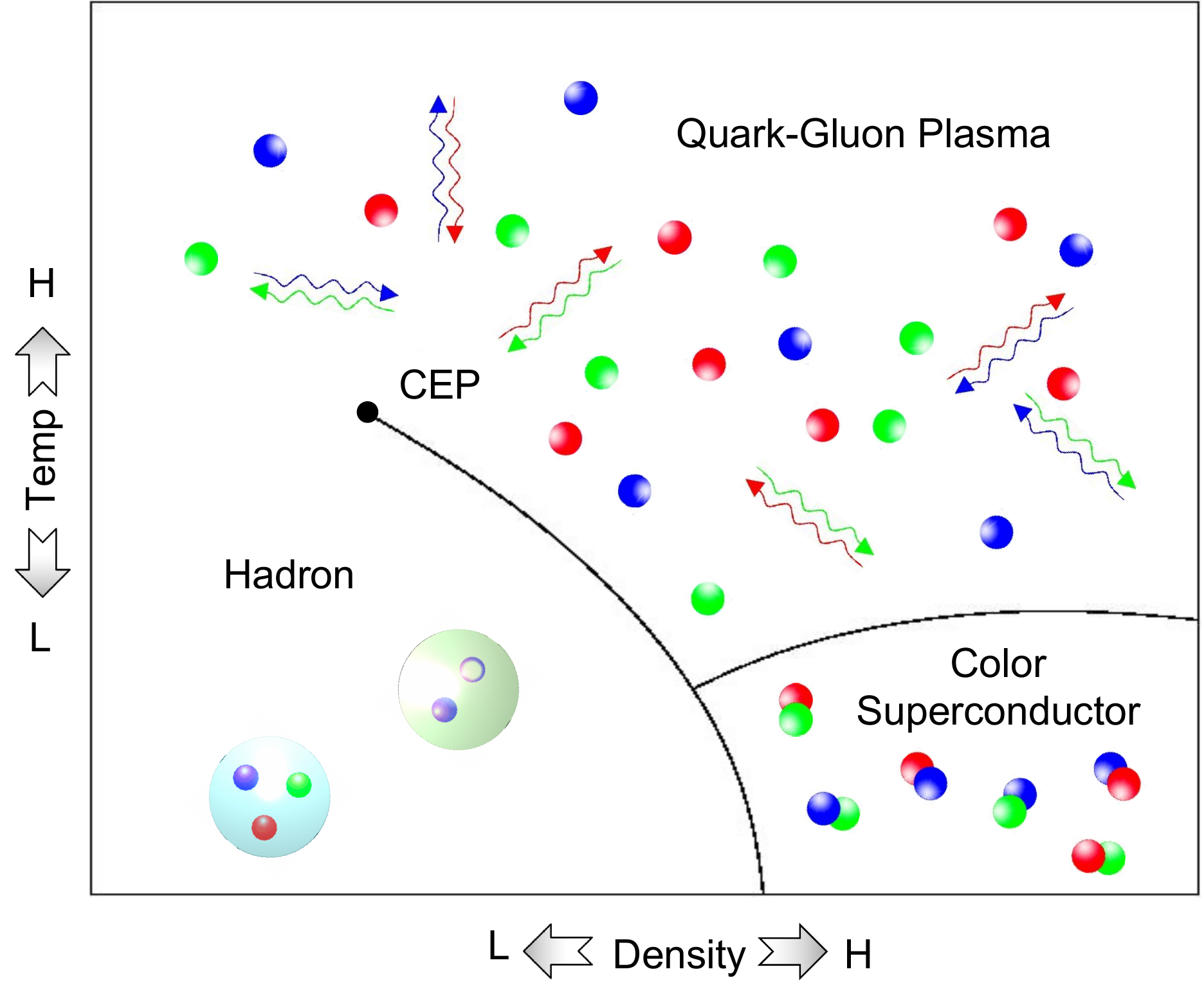}
    		\caption{Schematic view of expected phase diagram of QCD on the $T$-$\mu$ plane. As in Fig.~\ref{fig:phdgm_njl} the first-order phase transition line rises from the dense region at zero temperature and terminates at the critical end point (CEP).}
    		\label{fig:phdgm_qcd}
	\end{minipage}
\end{figure}

Recently, we have investigated the phase structure of the NJL model \cite{Nambu:1961tp,Nambu:1961fr} at extremely low temperature and high-density region on the lattice developing the Grassmann ATRG (GATRG) algorithm~\cite{Akiyama:2020soe}. The study of the NJL model has two important aspects. Firstly, the NJL model is a prototype of QCD. Their phase structures are expected to be similar as shown in Figs.~\ref{fig:phdgm_njl} and \ref{fig:phdgm_qcd}. The study of the NJL model at finite density is a good testbed before exploring the finite density QCD. Secondly, the NJL model has a similar path-integral form to the Hubbard model, a fundamentally important model to understand the strongly correlated electrons. Both models consist of hopping terms and a four-fermi interaction term. This fact indicates that the technical details of the TRG method employed in the analysis of the NJL model could be applied to the Hubbard model. Actually, we have analyzed the doping-driven metal-insulator transition of the (1+1)$d$ Hubbard model with the TRG method in Ref.~\cite{Akiyama:2021xxr} and our results for the critical chemical potential and the critical exponent show good consistency with the theoretical predictions based on the Bethe ansatz~\cite{PhysRevLett.20.1445,LIEB20031}. We have also extended this calculation to the (2+1)$d$ Hubbard model~\cite{Akiyama:2021glo}.

\subsection{TRG method for fermionic systems}
\label{subsec:gtrg}

There are several ways to introduce the tensor network representation for fermionic path integrals \cite{Shimizu:2014uva,Gu:2010yh,Takeda:2014vwa,Akiyama:2020sfo}. Here, we follow the formulation of Ref.~\cite{Akiyama:2020sfo}, where the fermionic path integrals are expressed by the Grassmann tensors. Let us now consider the following fermionic action as an example,
\begin{align}
	S[\bar{\psi},\psi]=\sum_{n\in\Lambda}\left[-t\sum_{\nu=1}^{d}\left(\bar{\psi}(n)\psi(n+\hat{\nu})+\bar{\psi}(n+\hat{\nu})\psi(n)\right)+W[\bar{\psi}(n),\psi(n)]\right].
\end{align}
$\psi(n)$ and $\bar{\psi}(n)$ are the fermion fields. For simplicity, we assume that they are single-component Grassmann fields. The path integral is
\begin{align}
	Z=\int\left(\prod_{n\in\Lambda}{\rm d}\bar{\psi}(n){\rm d}\psi(n)\right)~{\rm e}^{-S[\bar{\psi},\psi]}.
	\label{eq:gZ}
\end{align}
We decompose hopping factors introducing auxiliary Grassmann fields such that
\begin{align}
	{\rm e}^{t\bar{\psi}(n)\psi(n+\hat{\nu})}=\int{\rm d}\bar{\eta}_{\nu}(n){\rm d}\eta_{\nu}(n)~{\rm e}^{-\bar{\eta}_{\nu}(n)\eta_{\nu}(n)}~{\rm e}^{\sqrt{t}\bar{\psi}(n)\eta_{\nu}(n)}~{\rm e}^{-\sqrt{t}\psi(n+\hat{\nu})\bar{\eta}_{\nu}(n)},
\end{align}
\begin{align}
	{\rm e}^{t\bar{\psi}(n+\hat{\nu})\psi(n)}=\int{\rm d}\bar{\xi}_{\nu}(n){\rm d}\xi_{\nu}(n)~{\rm e}^{-\bar{\xi}_{\nu}(n)\xi_{\nu}(n)}~{\rm e}^{-\sqrt{t}\bar{\psi}(n+\hat{\nu})\bar{\xi}_{\nu}(n)}~{\rm e}^{-\sqrt{t}\psi(n)\xi_{\nu}(n)}.
\end{align}
Thanks to these decompositions, we are now allowed to integrate out $\psi(n)$ and $\bar{\psi}(n)$ independently at each site. The Grassmann tensor $\mathcal{T}$ is defined as a result of the integration,
\begin{align}
	\mathcal{T}=\int{\rm d}\bar{\psi}{\rm d}\psi~{\rm e}^{-W[\bar{\psi},\psi]}\prod_{\nu=1}^{d}{\rm e}^{\sqrt{t}\bar{\psi}\eta_{\nu}(n)}~{\rm e}^{-\sqrt{t}\psi\xi_{\nu}(n)}~{\rm e}^{-\sqrt{t}\bar{\psi}\bar{\xi}_{\nu}(n-\hat{\nu})}~{\rm e}^{-\sqrt{t}\psi\bar{\eta}_{\nu}(n-\hat{\nu})}.
	\label{eq:gtensor}
\end{align}
Since $(\eta_{\nu},\xi_{\nu})$ and $(\bar{\eta}_{\nu},\bar{\xi}_{\nu})$ play the roles of bond degrees of freedom, we regard them as subscripts of the Grassmann tensor; $\mathcal{T}_{\Psi_{1}\cdots\Psi_{d}\bar{\Psi}_{d}\cdots\bar{\Psi}_{1}}$ with $\Psi_{\nu}=(\eta_{\nu},\xi_{\nu})$ and $\bar{\Psi}_{\nu}=(\bar{\eta}_{\nu},\bar{\xi}_{\nu})$. The path integral of Eq.~\eqref{eq:gZ} is now expressed by
\begin{align}
	Z=\int\left(\prod_{n\in\Lambda}\prod_{\nu=1}^{d}{\rm d}\bar{\Psi}_{\nu}(n){\rm d}\Psi_{\nu}(n)~{\rm e}^{-\bar{\Psi}_{\nu}(n)\Psi_{\nu}(n)}\right)\prod_{n\in\Lambda}\mathcal{T}_{\Psi_{1}(n)\cdots\Psi_{d}(n)\bar{\Psi}_{d}(n-\hat{d})\cdots\bar{\Psi}_{1}(n-\hat{1})}.
	\label{eq:gtn}
\end{align}
We refer this expression as the Grassmann tensor network representation for $Z$. 

To apply a certain TRG algorithm to evaluate Eq.~\eqref{eq:gtn}, one needs to develop the corresponding algorithm extended to evaluate the Grassmann integral. To this aim, we rewrite Eq.~\eqref{eq:gtensor} in the following form,
\begin{align}
	\mathcal{T}=\left(\prod_{\nu=1}^{d}\sum_{i_{\nu},j_{\nu},i'_{\nu},j'_{\nu}}\right)T_{(i_{1}j_{1})\cdots(i_{d}j_{d})(i'_{1}j'_{1})\cdots(i'_{d}j'_{d})}\eta_{1}^{i_{1}}\xi_{1}^{j_{1}}\cdots\eta_{d}^{i_{d}}\xi_{d}^{j_{d}}\bar{\xi}_{d}^{j'_{d}}\bar{\eta}_{d}^{i'_{d}}\cdots\bar{\xi}_{1}^{j'_{1}}\bar{\eta}_{1}^{i'_{1}}.
	\label{eq:exp}
\end{align}
$T$ in the right hand side is a tensor in the usual sense.\footnote{One can easily obtain an explicit form of $T$, but it is not necessary in the following discussion.} In the practical TRG calculation, we have to encode the information of the Grassmann algebra into $T$ in some way. One of such ways is to encode the Grassmann parity for each $\Psi_{\nu}$, $\bar{\Psi}_{\nu}$ into the ordering of corresponding $\nu$-directional subscript in $T$. For example, let us map the $\nu$-directional subscript $(i_{\nu}j_{\nu})$ to the new one $I_{\nu}\in\mathbb{N}$ by
\begin{align}
	(00) \mapsto 1,~ (11) \mapsto 2,~ (10) \mapsto 3,~ (01) \mapsto 4.
\end{align}
Based on this mapping, we can regard $T$ in Eq.~\eqref{eq:exp} as a $2d$-rank tensor $T_{I_{1}\cdots I_{d}I'_{1}\cdots I'_{d}}$. For each subscript $I_{\nu}(I'_{\nu})$, the first two components correspond to the Grassmann-even sector of $\Psi_{\nu}(\bar{\Psi}_{\nu})$. When one implements the Grassmann TRG algorithm, it is necessary to read out the Grassmann parity from the subscript of $T$. This means that we need to define a binary function $f_{\nu}$ on $I_{\nu}$: $f_{\nu}(I_{\nu})=0(1)$ if $I_{\nu}$ corresponds to the Grassmann-even(odd) sector. Thanks to these binary functions, we can reproduce the Grassmann algebra just in $T$. For instance
\begin{align}
	T_{I_{1}I_{2}I_{3}\cdots I_{d}I'_{1}\cdots I'_{d}}=(-1)^{f_{1}(I_{1})f_{2}(I_{2})}T_{I_{2}I_{1}I_{3}\cdots I_{d}I'_{1}\cdots I'_{d}}
\end{align}
corresponds to the exchange between $\Psi_{1}$ and $\Psi_{2}$. The same argument also holds for the renormalized Grassmann tensor $\mathcal{T}'$ defined by a certain TRG algorithm, where the SVD,
\begin{align}
	Q_{abcd}\approx\sum_{k=1}^{D}U_{ab,k}\sigma_{k}V^{*}_{cd,k},
	\label{eq:svd}
\end{align}
is employed to introduce the coarse-grained degrees of freedom. Within the formulation explained above, each subscript has the information of the Grassmann parity, which allows us to consider the block-diagonal representation of Eq.~\eqref{eq:svd},
\begin{align}
	\begin{bmatrix}
		Q^{({\rm even})} & 0 \\
		0 & Q^{({\rm odd})}
	\end{bmatrix}
	\approx
	\begin{bmatrix}
		U^{({\rm even})} & 0 \\
		0 & U^{({\rm odd})}
	\end{bmatrix}
	\begin{bmatrix}
		\sigma^{({\rm even})} & 0 \\
		0 & \sigma^{({\rm odd})}
	\end{bmatrix}
	\begin{bmatrix}
		V^{({\rm even})\dag} & 0 \\
		0 & V^{({\rm odd})\dag}
	\end{bmatrix}
	.
\end{align}
In Eq.~\eqref{eq:svd}, the subscript $k$ corresponds to a new auxiliary Grassmann field in $\nu$-direction. In addition, if $\sigma_{k}$ belongs to $\sigma^{({\rm even})}(\sigma^{({\rm odd})})$, then $k$ represents the Grassmann-even(odd) component. In other words, the block-diagonalized SVD defines a new binary function $f_{\nu}$ for the coarse-grained auxiliary Grassmann field in $\nu$-direction. 

Now, it must be ready to extend a certain TRG algorithm to evaluate Eq.~\eqref{eq:gtn}. All we have to do is to carry out the TRG algorithm combining some phase factor $(-1)^{p}$ characterized by binary functions which reproduce the Grassmann algebra. In the following, we use the HOTRG \cite{PhysRevB.86.045139} or ATRG \cite{Adachi:2019paf} to evaluate fermionic path integrals. These algorithms consider a mapping like $\mathcal{T}\cdot\mathcal{T}\mapsto\mathcal{T}'$ along a certain direction. Suppose we make such a mapping along $\hat{1}$-direction, which firstly carries out the Grassmann integration,
\begin{align}
	&(\mathcal{TT})_{\Psi_{1}(n+\hat{1})\Xi_{2}\cdots\Xi_{d}\bar{\Xi}_{d}\cdots\bar{\Xi}_{2}\bar{\Psi}_{1}(n-\hat{1})}\nonumber\\
	&=\int{\rm d}\bar{\Psi}_{1}(n){\rm d}\Psi_{1}(n)~{\rm e}^{-\bar{\Psi}_{1}(n)\Psi_{1}(n)}
	\mathcal{T}_{\Psi_{1}(n+\hat{1})\cdots\Psi_{d}(n+\hat{1})\bar{\Psi}_{d}(n+\hat{1}-\hat{d})\cdots\bar{\Psi}_{1}(n)}
	\mathcal{T}_{\Psi_{1}(n)\cdots\Psi_{d}(n)\bar{\Psi}_{d}(n-\hat{d})\cdots\bar{\Psi}_{1}(n-\hat{1})},
	\label{eq:cg_g}
\end{align}
before we apply isometries (or squeezers) to accomplish the coarse-graining transformation $(\mathcal{TT})\mapsto\mathcal{T}'$. We have used shorthand notations defined by $\Xi_{\nu}=(\Psi_{\nu}(n+\hat{1})\Psi_{\nu}(n))$ and $\bar{\Xi}_{\nu}=(\bar{\Psi}_{\nu}(n-\hat{\nu})\bar{\Psi}_{\nu}(n+\hat{1}-\hat{\nu}))$. Therefore, introducing $\tilde{I}_{\nu}(n)=f_{\nu}(I_{\nu}(n))$, one can find
\begin{align}
\label{eq:g_sign}
	p&=\tilde{I}_{1}(n)\nonumber\\
	&+\tilde{I}_{2}(n)\tilde{I}_{2}(n+\hat{1})\nonumber\\
	&+\tilde{I}_{3}(n)[\tilde{I}_{2}(n+\hat{1})+\tilde{I}_{3}(n+\hat{1})]\nonumber\\
	&+\cdots\nonumber\\
	&+\tilde{I}_{d}(n)[\tilde{I}_{2}(n+\hat{1})+\cdots+\tilde{I}_{d}(n+\hat{1})]\nonumber\\
	&+\tilde{I}'_{d}(n-\hat{d})[\tilde{I}'_{d}(n+\hat{1}-\hat{d})+\cdots+\tilde{I}'_{2}(n+\hat{1}-\hat{2})]\nonumber\\
	&+\tilde{I}'_{d-1}(n-\widehat{d-1})[\tilde{I}'_{d-1}(n+\hat{1}-\widehat{d-1})+\cdots+\tilde{I}'_{2}(n+\hat{1}-\hat{2})]\nonumber\\
	&+\cdots\nonumber\\
	&+\tilde{I}'_{2}(n-\hat{2})\tilde{I}'_{2}(n+\hat{1}-\hat{2})
\end{align}
is the phase factor which makes the contraction
\begin{align}
	(TT)_{I_{1}(n+\hat{1})J_{2}\cdots J_{d}I'_{1}(n)J'_{2}\cdots J'_{d}}
	=\sum_{I_{1}(n)}(-1)^{p}T_{I_{1}(n+\hat{1})\cdots I_{d}(n+\hat{1})I_{1}(n)\cdots I'_{d}(n-\hat{d}+\hat{1})}T_{I_{1}(n)\cdots I_{d}(n)I'_{1}(n-\hat{1})\cdots I'_{d}(n-\hat{d})}
	\label{eq:cg_tensor}
\end{align}
equivalent to Eq.~\eqref{eq:cg_g}. Note that we have introduced shorthand notations $J_{\nu}=(I_{\nu}(n+\hat{1})I_{\nu}(n))$ and $J'_{\nu}=(I'_{\nu}(n+\hat{1}-\hat{\nu})I'_{\nu}(n-\hat{\nu}))$. It is a very straightforward task to develop the Grassmann version of the HOTRG or ATRG (or also the triad RG \cite{Kadoh:2019kqk}) reflecting on Eq.~\eqref{eq:cg_tensor}. \footnote{When one assumes the anti-periodic boundary condition in $\hat{\nu}$-direction, an additional phase factor $(-1)^{\tilde{I}_{\nu}}$ is necessary just in taking the trace of $T$.}

\subsection{(3+1)$d$ NJL model on the lattice}
\label{subsec:njl}

The Lagrangian of the NJL model in the continuum is defined as follows:
\begin{align}
	\label{eq:njl_cont}
	{\cal L}=\sum_{\nu=1}^{4}{\bar \psi}(x)\gamma_\nu\partial_\nu \psi(x) 
 	-g_0\left\{({\bar \psi}(x)\psi(x))^2+({\bar \psi}(x){\rm i}\gamma_5\psi(x))^2 \right\},
\end{align}
which has the U(1) chiral symmetry with $\psi(x) \rightarrow {\rm e}^{{\rm i}\alpha\gamma_5}\psi(x)$ and ${\bar \psi}(x) \rightarrow {\bar \psi}(x){\rm e}^{{\rm i}\alpha\gamma_5}$. A schematic view of the expected phase structure on the $T$-$\mu$ plane is depicted in Fig.~\ref{fig:phdgm_njl}, where a characteristic feature is the first-order chiral phase transition in the dense region at very low temperature~\cite{Asakawa:1989bq}. We have investigated the phase transition employing the chiral condensate as an order parameter with the Kogut-Susskind fermion to formulate the NJL model on  the lattice. Following Refs.~\cite{Lee:1987eg,Booth:1989ms}, we define the model at finite chemical potential $\mu$ as
\begin{align}
	\label{eq:njl_lat}
	S=&\frac{1}{2}a^3\sum_{n\in\Lambda}\sum_{\nu=1}^{4} \eta_\nu(n)\left[{\rm e}^{\mu a\delta_{\nu,4}}{\bar \chi}(n)\chi(n+{\hat\nu})-{\rm e}^{-\mu a\delta_{\nu,4}}{\bar \chi}(n+{\hat\nu})\chi(n)\right]\nonumber\\
	&+ma^4\sum_{n\in\Lambda}{\bar \chi}(n)\chi(n)-g_{0}a^4\sum_{n\in\Lambda}\sum_{\nu=1}^{4}{\bar \chi}(n)\chi(n){\bar \chi}(n+{\hat\nu})\chi(n+{\hat\nu}),
\end{align}
where $n=(n_1,n_2,n_3,n_4)\in\Lambda(\subset\mathbb{Z}^4)$ specifies a position in lattice $\Lambda$ with the lattice spacing $a$. $\chi(n)$ and $\bar{\chi}(n)$ are Grassmann-valued fields without the Dirac structure. Since they describe the Kogut-Susskind fermions, $\chi(n)$ and $\bar{\chi}(n)$ are single-component Grassmann variables. $\eta_{\nu}(n)$ is the staggered sign function defined by $\eta_\nu(n)=(-1)^{n_1+\cdots +n_{\nu-1}}$ with $\eta_1(n)=1$.
The four-fermi coupling is chosen to be $g_0=32$.
The path integral is defined in an ordinary manner:
\begin{align}
\label{eq:partition}
	Z=\int\left(\prod_{n\in\Lambda}{\rm d}\chi(n){\rm d}\bar{\chi}(n)\right){\rm e}^{-S}.
\end{align}
For vanishing mass $m\rightarrow 0$, Eq.~\eqref{eq:njl_lat} is invariant under the following continuous chiral transformation:
\begin{align}
	\chi(n)&\rightarrow {\rm e}^{{\rm i}\alpha\epsilon(n)}\chi(n),\\
	{\bar \chi}(n)&\rightarrow {\bar \chi}(n){\rm e}^{{\rm i}\alpha\epsilon(n)}
\end{align}
with $\alpha\in\mathbb{R}$ and $\epsilon(n)=(-1)^{n_1+n_2+n_3+n_4}$. 

After rewriting the path integral in the tensor network representation, we evaluate it using the GATRG algorithm on a lattice up to the volume of $V=L^4$ ($L=2^m, m \in \mathbb{N}$). The technical details for the tensor network representation and the GATRG procedure are given in Ref.~\cite{Akiyama:2020soe}. We employ the periodic boundary conditions for $x$-, $y$-, $z$-directions and the anti-periodic boundary condition for $t$-direction.

\subsubsection{Heavy dense limit as a benchmark}

We first check the efficiency of the GATRG algorithm by benchmarking with the NJL model in the heavy dense limit, which is defined as $m\to\infty$ and $\mu\to\infty$ with ${\rm e}^{\mu}/m$ kept fixed. The heavy dense limit gives us an opportunity to compare numerical results with the exact analytical ones, whose expressions for the number density $\langle n\rangle$ and the fermion condensate $\langle {\bar \chi}(n)\chi(n)\rangle$ at vanishing temperature are given by the step functions
\begin{align}
\label{eq:hd_n}
	 \langle n\rangle=\Theta(\mu-\mu_{\rm c}),
\end{align}
\begin{align}
\label{eq:hd_cc}
	 \langle{\bar \chi}(n)\chi(n)\rangle=\frac{1}{m}\Theta(\mu-\mu_{\rm c}),
\end{align}
with $\mu_{\rm c}=\ln(2m)$ \cite{Pawlowski:2013gag}. 

\begin{figure}[htbp]
	\begin{minipage}[t]{0.48\hsize}
    		\centering
   		 \includegraphics[width=1.0\hsize]{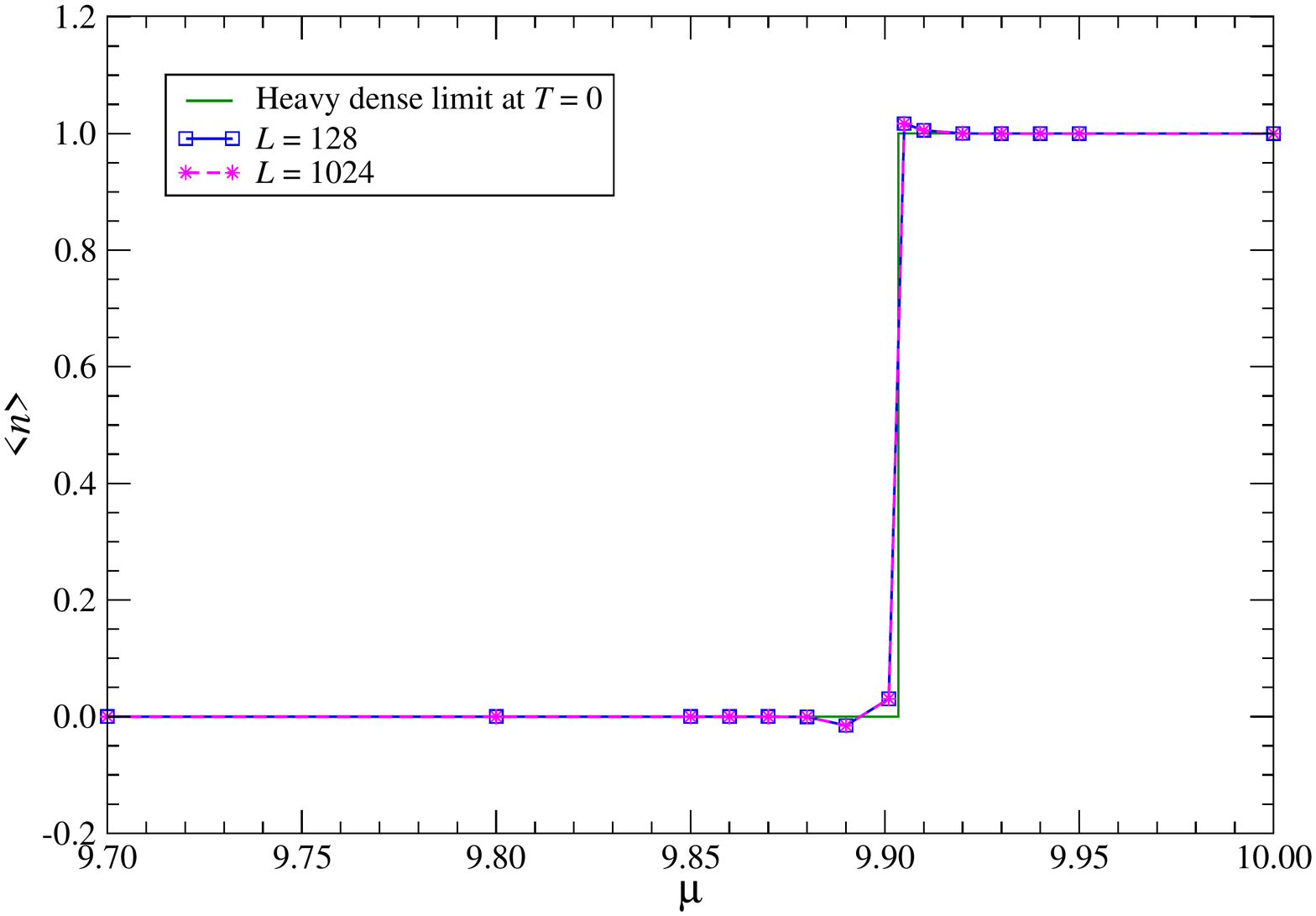}
   	 	\caption{Number density at $m=10^{4}$ and $g_0=32$ on $128^4$ and $1024^4$ lattices as a function of $\mu$ with $D=30$. $\Delta\mu=4.0\times10^{-3}$ in the vicinity of $\mu_{\rm c}$. Green line denotes the step function in Eq.~\eqref{eq:hd_n}.}
    		\label{fig:hd_n}
  	\end{minipage}
  	\hspace*{3mm}
	\begin{minipage}[t]{0.48\hsize}
    		\centering
    		\includegraphics[width=1.0\hsize]{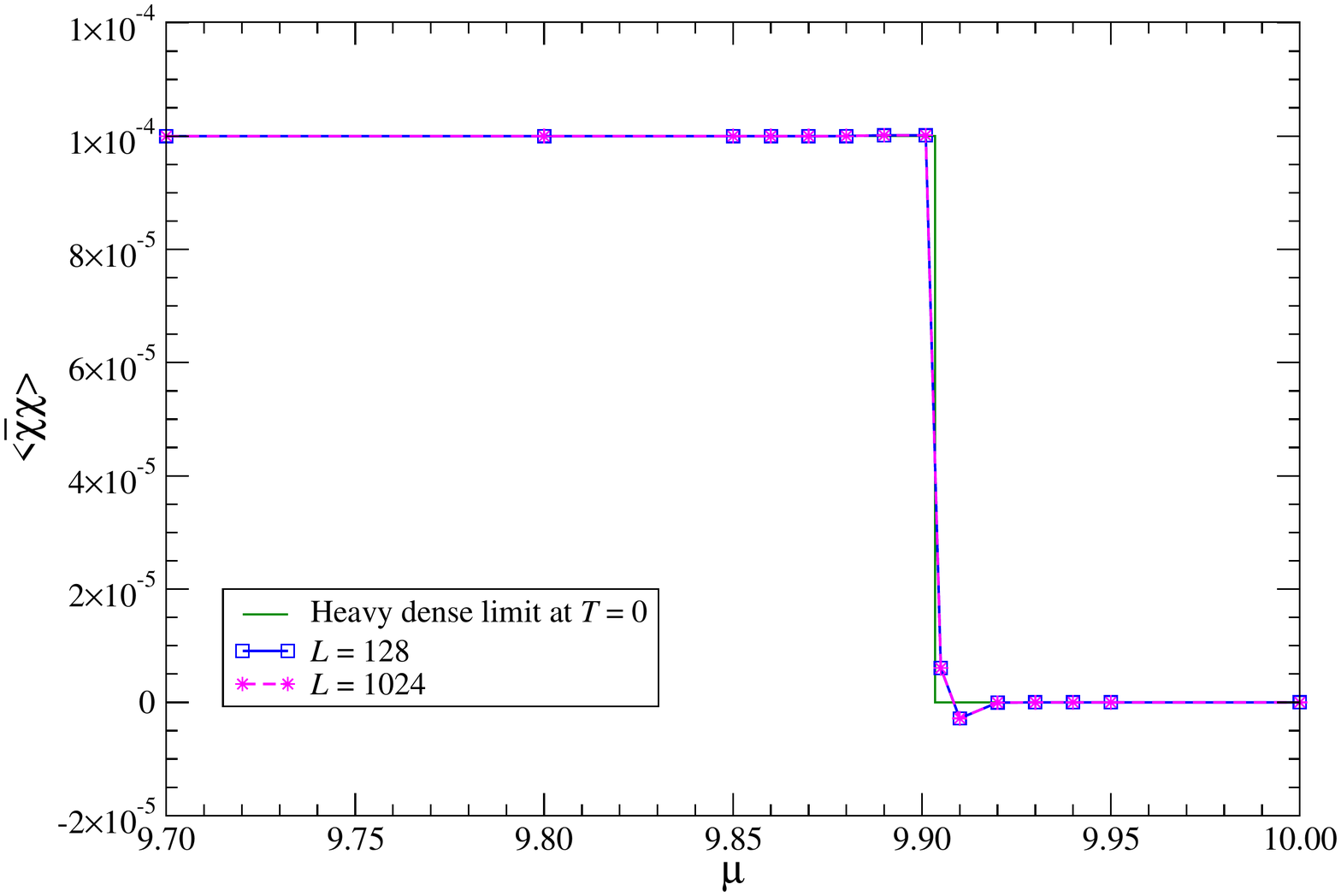}
   	 	\caption{Fermion condensate at $m=10^{4}$ and $g_0=32$ on $128^4$ and $1024^4$ lattices as a function of $\mu$ with $D=30$. Green line denotes the step function in Eq.~\eqref{eq:hd_cc}.}
    		\label{fig:hd_cc}
	\end{minipage}
\end{figure}

Figures~\ref{fig:hd_n} and \ref{fig:hd_cc} show the numerical results for $\langle n\rangle$ and $\langle {\bar \chi}(n)\chi(n)\rangle$ obtained by the GATRG algorithm choosing $m=10^4$ with the bond dimension $D=30$. 
The number density is calculated by the numerical derivative of the thermodynamic potential in terms of the chemical potential:
\begin{align}
\label{eq:nd_number}
	\langle n\rangle=\frac{1}{V}\frac{\partial \ln Z(\mu)}{\partial\mu}\approx\frac{1}{V}\frac{\ln Z(\mu+\Delta \mu)-\ln Z(\mu)}{\Delta \mu}.
\end{align}
In the vicinity of $\mu_{\rm c}$, we have set $\Delta\mu=4.0\times10^{-3}$. The fermion condensate is also obtained via the numerical derivative of the thermodynamic potential in terms of $m$:
\begin{align}
	\left.\langle{\bar \chi}(n)\chi(n)\rangle\right|_{m=10^4}=\left.\frac{1}{V}\frac{\ln Z(m+\Delta m)-\ln Z(m)}{\Delta m}\right|_{m=10^4}
\end{align}
with $\Delta m=1$. 
Since there is little difference between the $L=128$ and $1024$ results, the $L=1024$ lattice is sufficiently large to be estimated as the thermodynamic limit at vanishing temperature. The numerical results well reproduce the analytical ones, including the location of $\mu_{\rm c}=\ln(2m)=9.903$, both for $\langle n\rangle$ and $\langle {\bar \chi}(n)\chi(n)\rangle$ in the heavy dense limit.

\subsubsection{Chiral phase transition}

\begin{figure}[htbp]
	\begin{minipage}[t]{0.48\hsize}
    		\centering
    		\includegraphics[width=1.0\hsize]{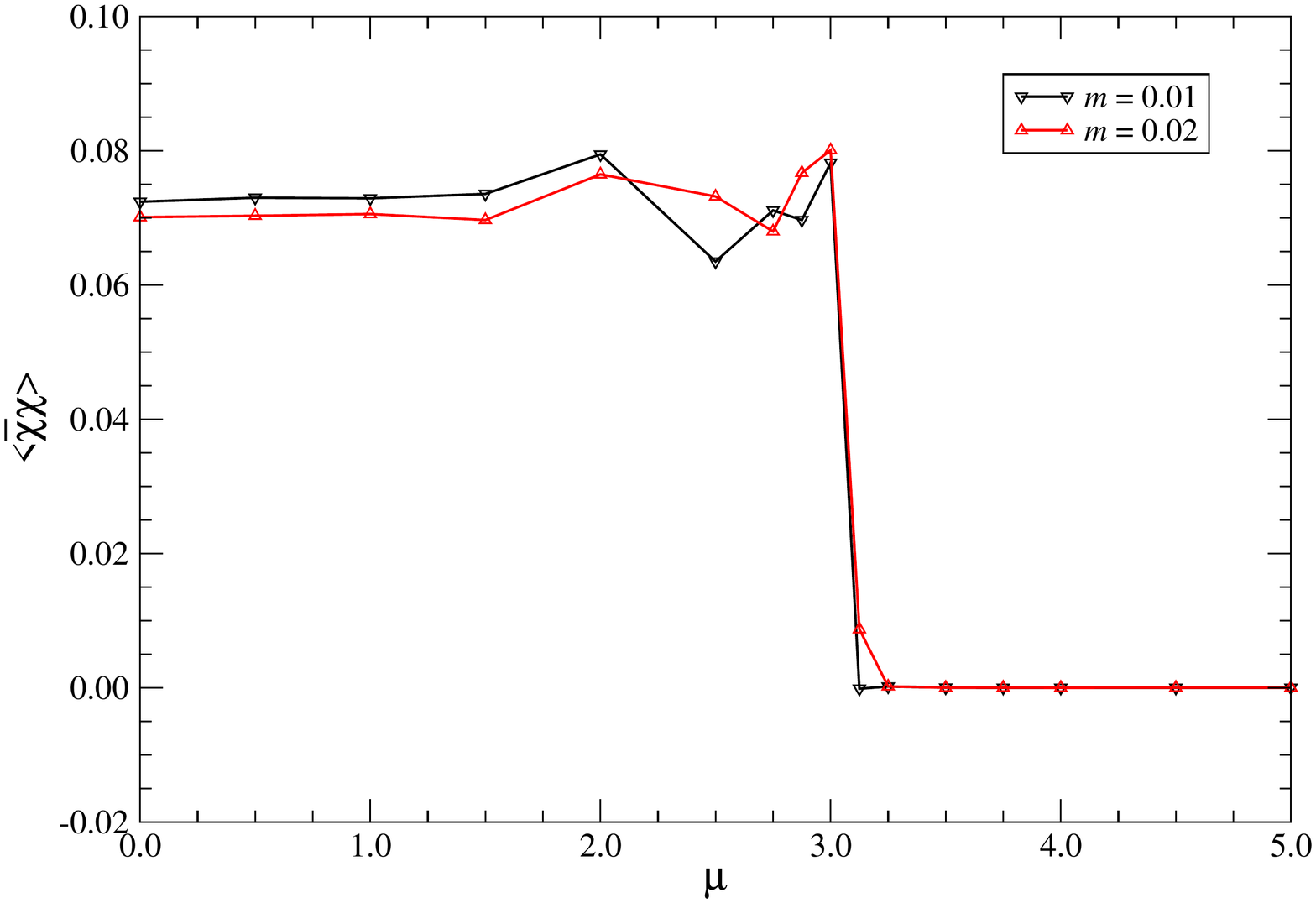}
  		\caption{Chiral condensate at $m=0.01$ and $0.02$ on $1024^4$ lattice as a function of $\mu$ with $D=55$.}
  		\label{fig:cc}
  	\end{minipage}
  	\hspace*{3mm}
	\begin{minipage}[t]{0.48\hsize}
    		\centering
   		 \includegraphics[width=1.0\hsize]{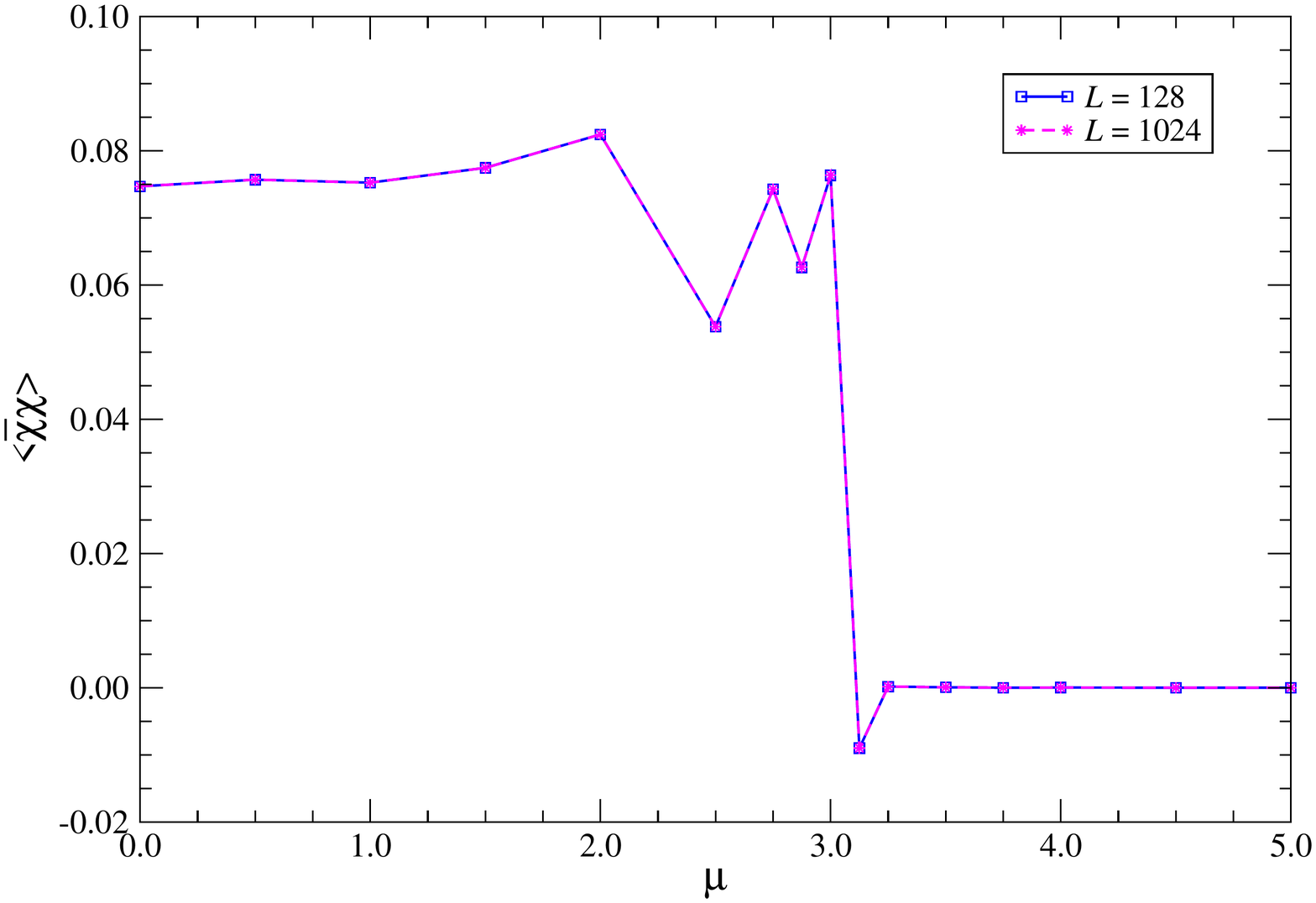}
  			\caption{Chiral condensate extrapolated in the chiral limit as a function of $\mu$ with $D=55$ on $128^4$ and $1024^4$ lattices.}
  		\label{fig:cc_m0}
	\end{minipage}
\end{figure}

The chiral condensate $\langle {\bar \chi}(n)\chi(n)\rangle$, which is an order parameter in the chiral phase transition, is defined by
  \begin{align}
\label{eq:def_cc}
	\langle {\bar \chi}(n)\chi(n)\rangle=\lim_{m\to0}\lim_{V\to\infty}\frac{1}{V}\frac{\partial}{\partial m}\ln Z,
\end{align}
  in the cold region. We calculate $\langle {\bar \chi}(n)\chi(n)\rangle$ with the numerical derivative of thermodynamic potential at $m=0.01$ and 0.02 and their chiral extrapolation in the thermodynamic limit. The numerical derivative is performed as 
  \begin{align}
	\frac{\partial}{\partial m}\ln Z\approx\frac{\ln Z(m+\Delta m)-\ln Z(m)}{\Delta m},
\end{align}
with $\Delta m=0.01$. Figure~\ref{fig:cc} shows the $\mu$ dependence of the chiral condensate at $m=0.01$ and 0.02 on the $V=1024^4$ lattice. The signals show slight fluctuations as a function of $\mu$ around the transition point.  Away from the transition point, we have found little response in $\langle {\bar \chi}(n)\chi(n)\rangle$ to changes in mass. Figure~\ref{fig:cc_m0} presents the results in the chiral limit obtained by the chiral extrapolation with the data at $m=0.01$ and $0.02$ on two volumes of $V=128^4$ and $1024^4$. The little discrepancy between the $L=128$ and $1024$ results means that the $L=1024$ result is essentially in the thermodynamic limit. We observe the discontinuity from a finite value to zero for the chiral condensate at $\mu_{\rm c}= 3.0625\pm 0.0625$, which is a clear indication of the first-order phase transition.

\subsubsection{Equation of state}

\begin{figure}[htbp]
	\begin{minipage}[t]{0.48\hsize}
    		\centering
   		 \includegraphics[width=1.0\hsize]{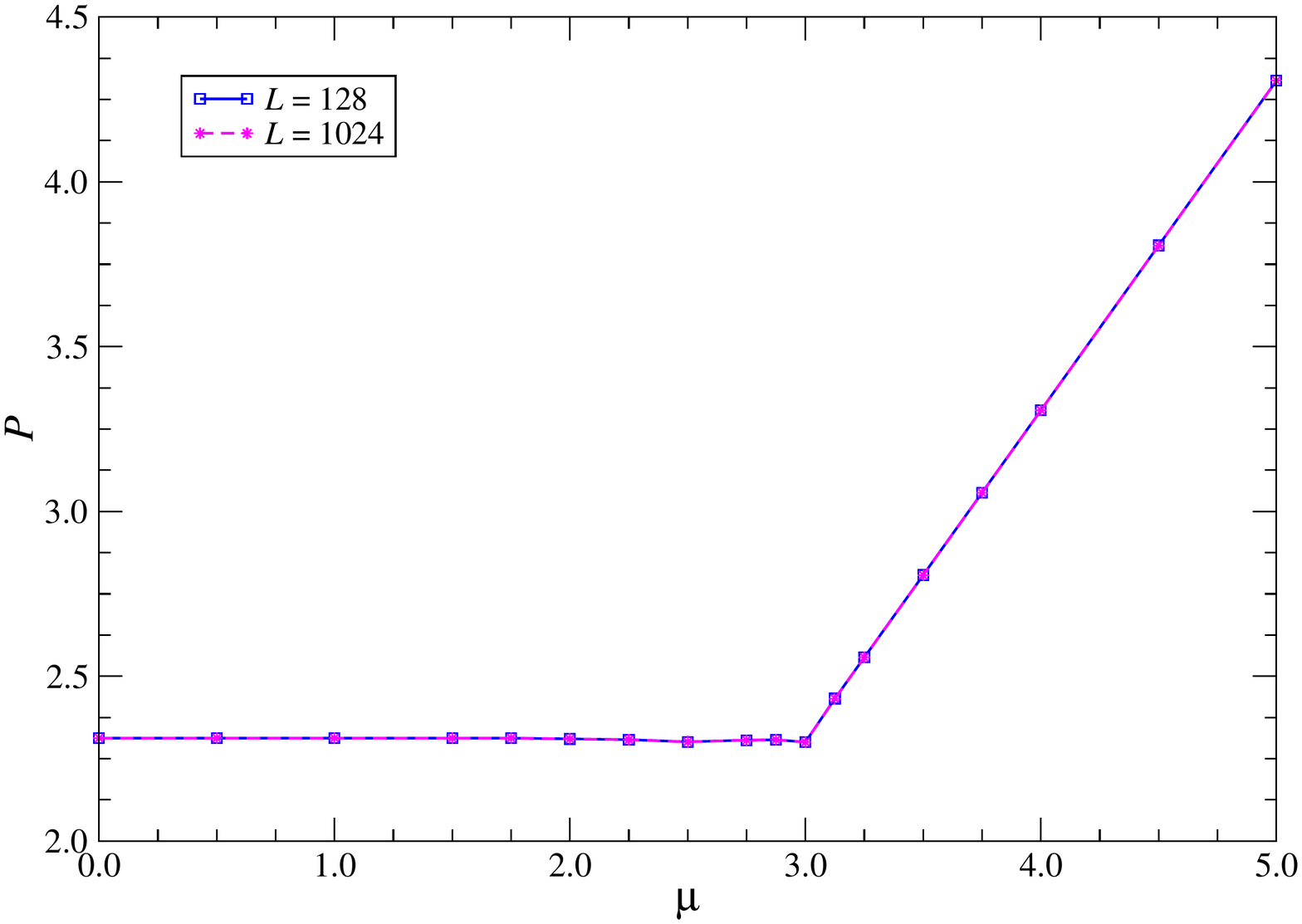}
   		\caption{Pressure at $m=0.01$ as a function of $\mu$ on $128^4$ and $1024^4$ lattices.}
  		\label{fig:pressure}
 	\end{minipage}
  	\hspace*{3mm}
	\begin{minipage}[t]{0.48\hsize}
    		\centering
   	 	\includegraphics[width=1.0\hsize]{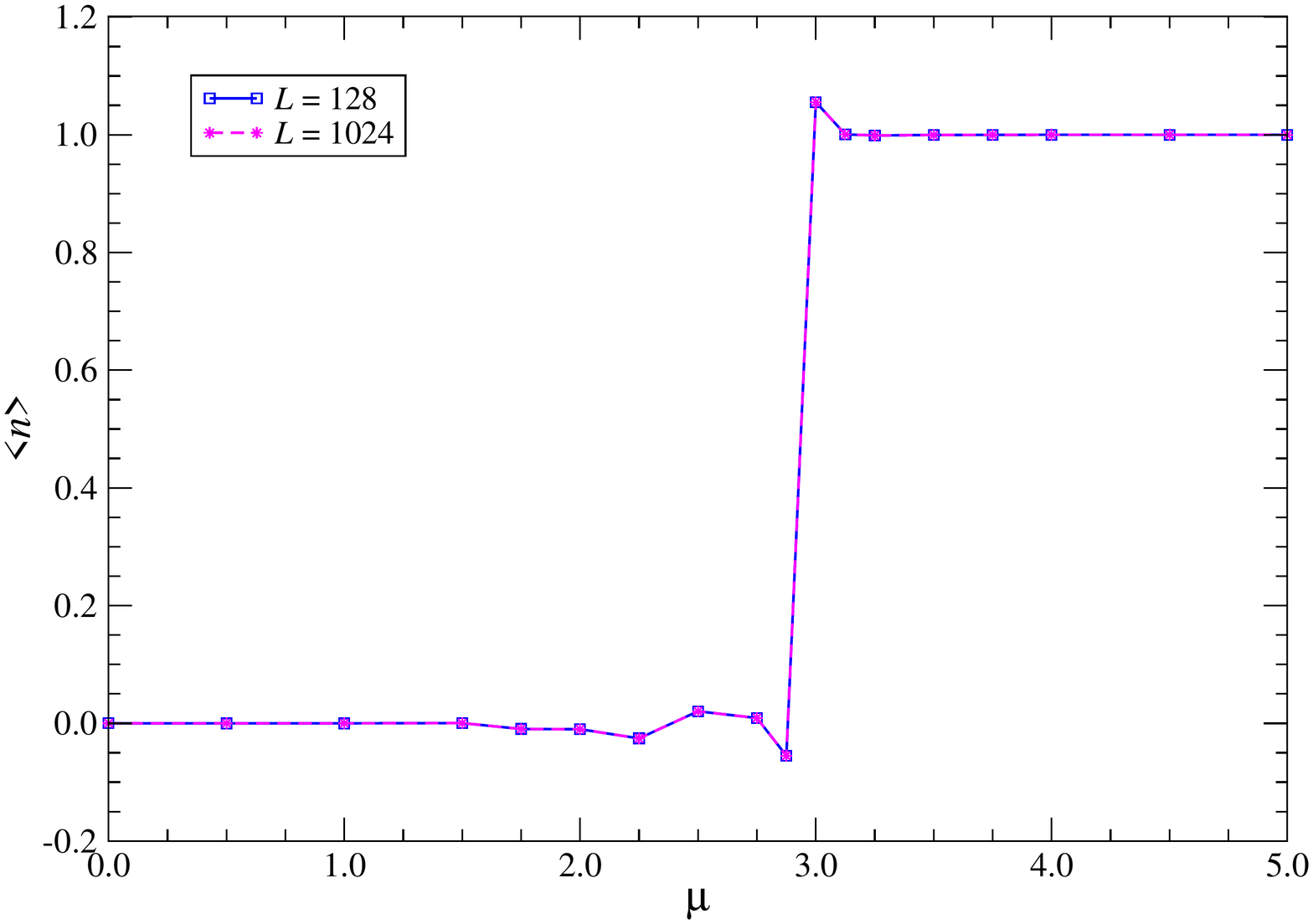}
  		\caption{Particle number density at $m=0.01$ as a function of $\mu$ on $128^4$ and $1024^4$ lattices.}
  		\label{fig:number_density}
	\end{minipage}
\end{figure}

Equation of state is a relation between the pressure and the particle number density. 
In the thermodynamic limit, the pressure $P$ is directly obtained from the thermodynamic potential:
\begin{align}
	P=\frac{\ln Z}{V},
\end{align}
where the vast homogeneous system is assumed. In Fig.~\ref{fig:pressure}, we plot the $\mu$ dependence of the pressure at $m=0.01$. We find a kink behavior at $\mu_{\rm c}= 3.0625\pm 0.0625$, where the chiral condensate shows the discontinuity. Note that the $m=0.02$ result shows little difference from the $m=0.01$ one.
Figure~\ref{fig:number_density} shows the $\mu$ dependence of the particle number density $\langle n\rangle$  obtained by Eq.~\eqref{eq:nd_number}. We observe an abrupt jump from $\langle n\rangle=0$ to $\langle n\rangle=1$ at $\mu_{\rm c}= 2.9375\pm 0.0625$. This is another indication of the first-order phase transition. 

\subsection{(1+1)$d$ Hubbard model}
\label{subsec:hubbard_1+1}

The Hubbard model has a similar path-integral form to the NJL model. The successful analysis of the phase transition of the (3+1)$d$ NJL model at high density and vanishing temperature with the TRG method urges us to apply it to investigate the metal-insulator transition of the (1+1)$d$ Hubbard model. Calculating the electron density as a function of the chemical potential $\mu$, we have determined the critical value of the chemical potential $\mu_{\rm c}$ and the critical exponent $\nu$ in the thermodynamic limit at zero temperature and compare them with an exact solution based on the Bethe ansatz~\cite{PhysRevLett.20.1445,LIEB20031}.

\subsubsection{Formulation and numerical algorithm}
\label{subsubsec:hubbard_1+1_formulation}

For later convenience, we consider the partition function of the Hubbard model on the $(d+1)$-dimensional anisotropic lattice with the physical volume $V=L^d\times \beta$, whose spatial extension is defined as $L=aN_{\sigma}$ with $a$ the spatial lattice spacing and $\sigma$ labels the spatial direction. $\beta$ denotes the inverse temperature, which is divided as $\beta=1/T=\epsilon N_\tau$. The path-integral expression of the partition function is given by  
\begin{align}
	Z=\int\left(\prod_{n\in\Lambda_{d+1}}\prod_{s=\uparrow,\downarrow}{\rm d}\bar{\psi}_{s}(n){\rm d}\psi_{s}(n)\right){\rm e}^{-S},
	\label{eq:Z}
\end{align}
where $n=((n_{\sigma})_{\sigma=1,\cdots,d},n_{\tau})\in\Lambda_{d+1}(\subset\mathbb{Z}^{d+1})$ specifies a position in the $(d+1)$-dimensional lattice. Since the Hubbard model describes the spin-1/2 fermions, they are labeled by $s=\uparrow,\downarrow$, corresponding to the spin-up and spin-down, respectively. Introducing the notation,
\begin{align}
	\psi(n)=\left(
	\begin{array}{c}
	 	\psi_\uparrow(n)\\ \psi_\downarrow(n) 
	\end{array}
	\right),
	~\bar{\psi}(n)=\left(\bar{\psi}_\uparrow(n),\bar{\psi}_\downarrow(n)\right),
\end{align}
the action $S$ is given by\footnote{See Ref.~\cite{Creutz:1986ky} or Refs.~\cite{10.2307/2033649,Suzuki:1976be} for the conversion procedure from the operator formalism to the path-integral one.}
\begin{align}
\label{eq:d+1_action}
        S&=\sum_{n\in\Lambda_{d+1}}\epsilon\left\{\bar{\psi}(n)\left(\frac{\psi(n+{\hat \tau})-\psi(n)}{\epsilon}\right)\right.\nonumber\\
        &\quad\left.-t\sum_{\sigma=1}^{d}\left(\bar{\psi}(n+{\hat\sigma})\psi(n)+\bar{\psi}(n)\psi(n+{\hat\sigma})\right)
         +\frac{U}{2}\left(\bar{\psi}(n)\psi(n)\right)^2-\mu\bar{\psi}(n)\psi(n)\right\}.
\end{align}
The choice of $d=1$ corresponds to the $(1+1)$-dimensional case.
The kinetic term in the spatial direction contains the hopping parameter $t$. The four-fermi interaction term represents the Coulomb repulsion of electrons at the same lattice site. In addition to the target parameter set of $(U,t)=(4,1)$, we consider two simplified cases of $(U,t)=(4,0)$ and (1,0) as a bench mark: The former is the atomic limit and the latter represents the free electrons. The chemical potential is denoted by the parameter $\mu$. Note that the half-filling is realized at $\mu=U/2$ in the current definition. We assume the periodic boundary condition in the spatial direction, $\psi(N_{\sigma}+1,n_\tau)=\psi(1,n_\tau)$, while the anti-periodic one in the temporal direction, $\psi(n_{\sigma},N_\tau+1)=-\psi(n_{\sigma},1)$. In the following discussion, we always set $a=1$.

We employ the HOTRG algorithm \cite{PhysRevB.86.045139} to evaluate the Grassmann tensor network representation of Eq.~(\ref{eq:Z}), whose derivation is given in Ref.~\cite{Akiyama:2021xxr}. Using the HOTRG, we firstly carry out $m_{\tau}$ times of renormalization along the temporal direction. This procedure converts the initial Grassmann tensor $\mathcal{T}_{\Psi_{\sigma}\Psi_{\tau}\bar{\Psi}_{\tau}\bar{\Psi}_{\sigma}}$ into the coarse-grained one $\mathcal{T}'_{\Xi_{\sigma}\Psi_{\tau}\bar{\Psi}_{\tau}\bar{\Xi}_{\sigma}}$. Secondly, we employ the $2d$ HOTRG procedure, regarding $\mathcal{T}'_{\Xi_{\sigma}\Psi_{\tau}\bar{\Psi}_{\tau}\bar{\Xi}_{\sigma}}$ as the initial tensor, to obtain the coarse-grained Grassmann tensor $\mathcal{T}''_{\Xi'_{\sigma}\Psi'_{\tau}\bar{\Psi}'_{\tau}\bar{\Xi}'_{\sigma}}$. Note that with sufficiently small $\epsilon(<1)$, little truncation error is accumulated with the first $m_{\tau}$ times of renormalization along $\tau$-direction. This is because the contribution from the spatial hopping terms of $O(\epsilon)$ is smaller than that from the temporal one of $O(1)$. For the $(1+1)d$ Hubbard model, we found that the optimal $m_{\tau}$ satisfied the condition $\epsilon 2^{m_{\tau}}\sim O(10^{-1})$. 

\subsubsection{$(U,t)=(4,0)$ and $(0,1)$ cases as a benchmark}

\begin{figure}[htbp]
	\begin{minipage}[t]{0.48\hsize}
    		\centering
    		\includegraphics[width=1.0\hsize]{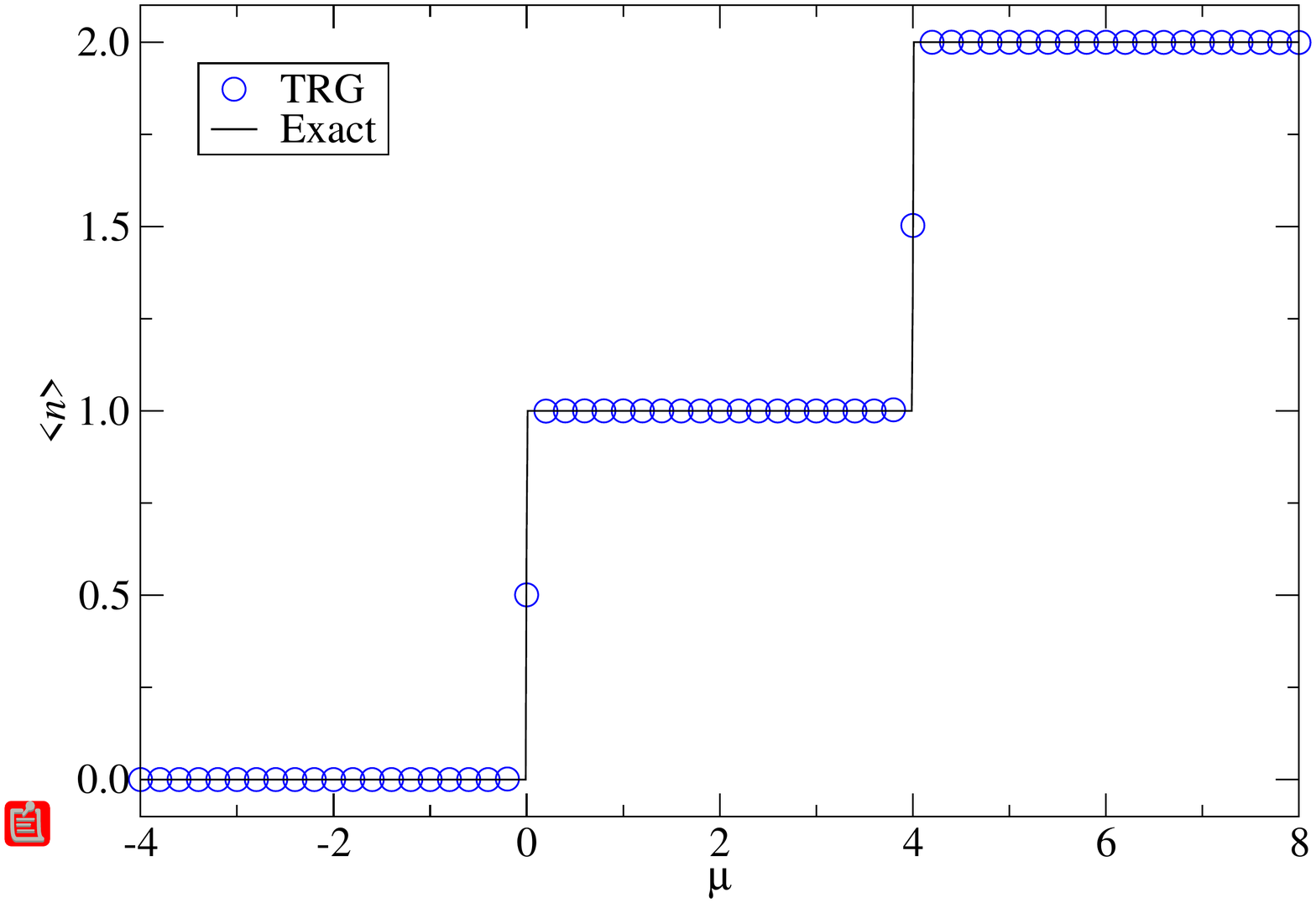}
		\caption{Electron density $\langle n\rangle$ in the $(U,t)=(4,0)$ case at $\beta=1677.7216$ with $\epsilon=10^{-4}$ as a function of $\mu$. The solid line shows the exact solution and the blue circles are the results obtained by the TRG method.}
  		\label{fig:edensity_t0}
  	\end{minipage}
  	\hspace*{3mm}
	\begin{minipage}[t]{0.48\hsize}
    		\centering
    		\includegraphics[width=1.0\hsize]{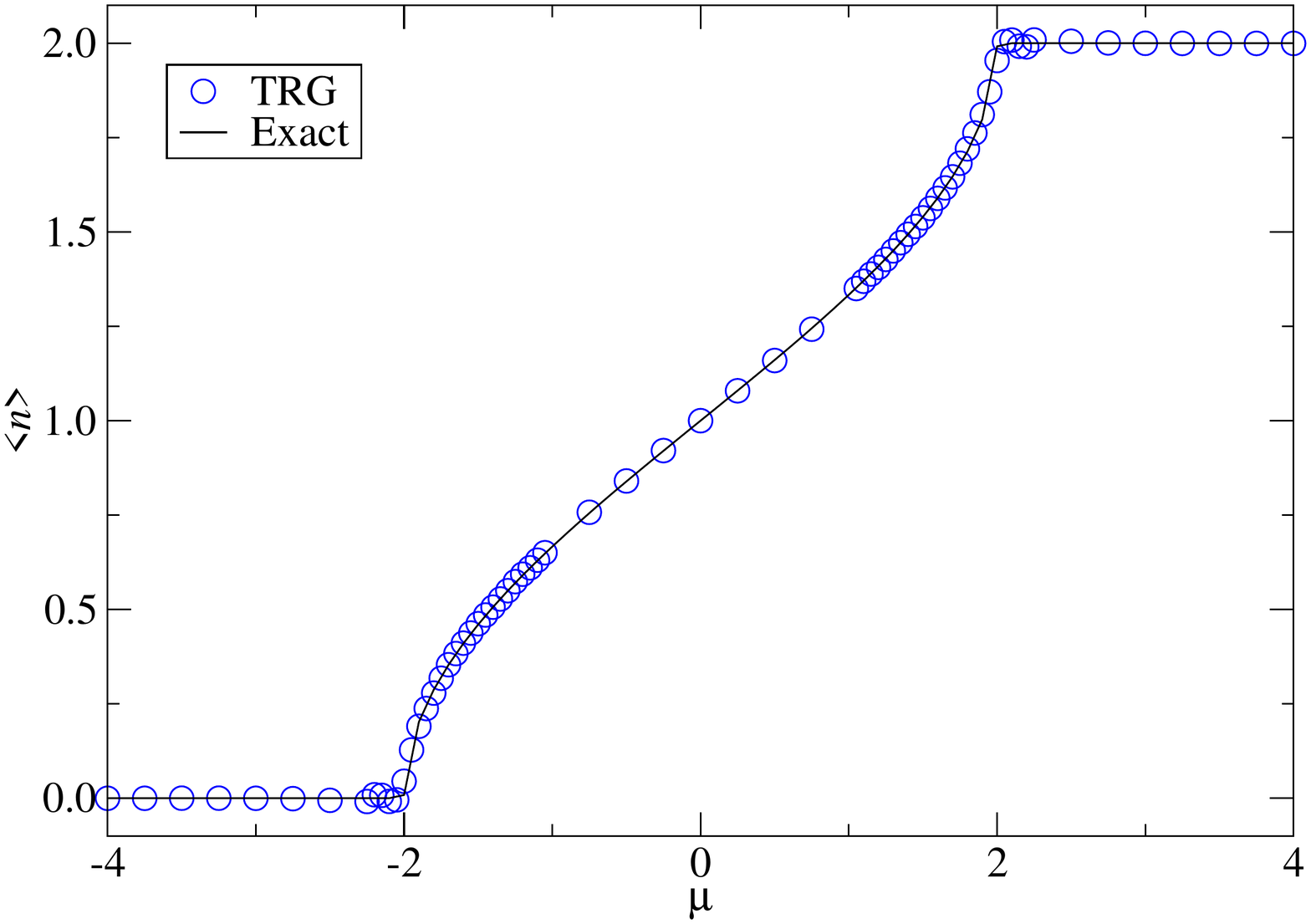}
		\caption{Electron density $\langle n\rangle$ in the $(U,t)=(0,1)$ case at $N_{\sigma}=4096$ and $\beta=1677.7216$ with $\epsilon=10^{-4}$ as a function of $\mu$. The solid line shows the exact solution on $N_{\sigma}=4096$ and the blue circles are the results obtained by the TRG method with $D=80$.}
  		\label{fig:edensity_U0}
	\end{minipage}
\end{figure}

We compare the numerical and analytic results for the $\mu$ dependence of $\langle n\rangle$ in two extreme cases of $(U,t)=(4,0)$ and (0,1).
The electron density $\langle n\rangle$ is obtained by the numerical derivative of the thermodynamic potential in terms of $\mu$:
\begin{align}
	\langle n\rangle=\frac{1}{V}\frac{\partial \ln Z(\mu)}{\partial \mu}\approx
	\frac{1}{V}\frac{\ln Z(\mu+\Delta \mu)-\ln Z(\mu-\Delta \mu)}{2\Delta \mu}.
        \label{eq:nd_hubbard}
\end{align}
We choose $\epsilon=10^{-4}$ for the discretization parameter in the temporal direction and $D=80$ for the truncation parameter after investigating the $\epsilon$ and $D$ dependences of the free energy systematically. In Figs.~\ref{fig:edensity_t0} and \ref{fig:edensity_U0} the numerical and analytic results show good consistencies over the wide range of $\mu$ in both cases. Note that for the case of $(U,t)=(4,0)$ in Fig.~\ref{fig:edensity_t0}, we set $m_{\tau}=24$ because this case is equivalent to the model defined on $V=1\times\beta$ lattice. Thanks to the vanishing hopping structure in the spatial direction, we can always perform an exact tensor contraction in the temporal direction. In Fig.~\ref{fig:edensity_U0} we employ finer resolution of $\mu$ around $1\lesssim|\mu|\lesssim2$ in order to follow the complicated $\mu$ dependence of $\langle n\rangle$.

\subsubsection{$(U,t)=(4,1)$ case}

\begin{figure}[htbp]
	\begin{minipage}[t]{0.48\hsize}
   		 \centering
    		\includegraphics[width=1.0\hsize]{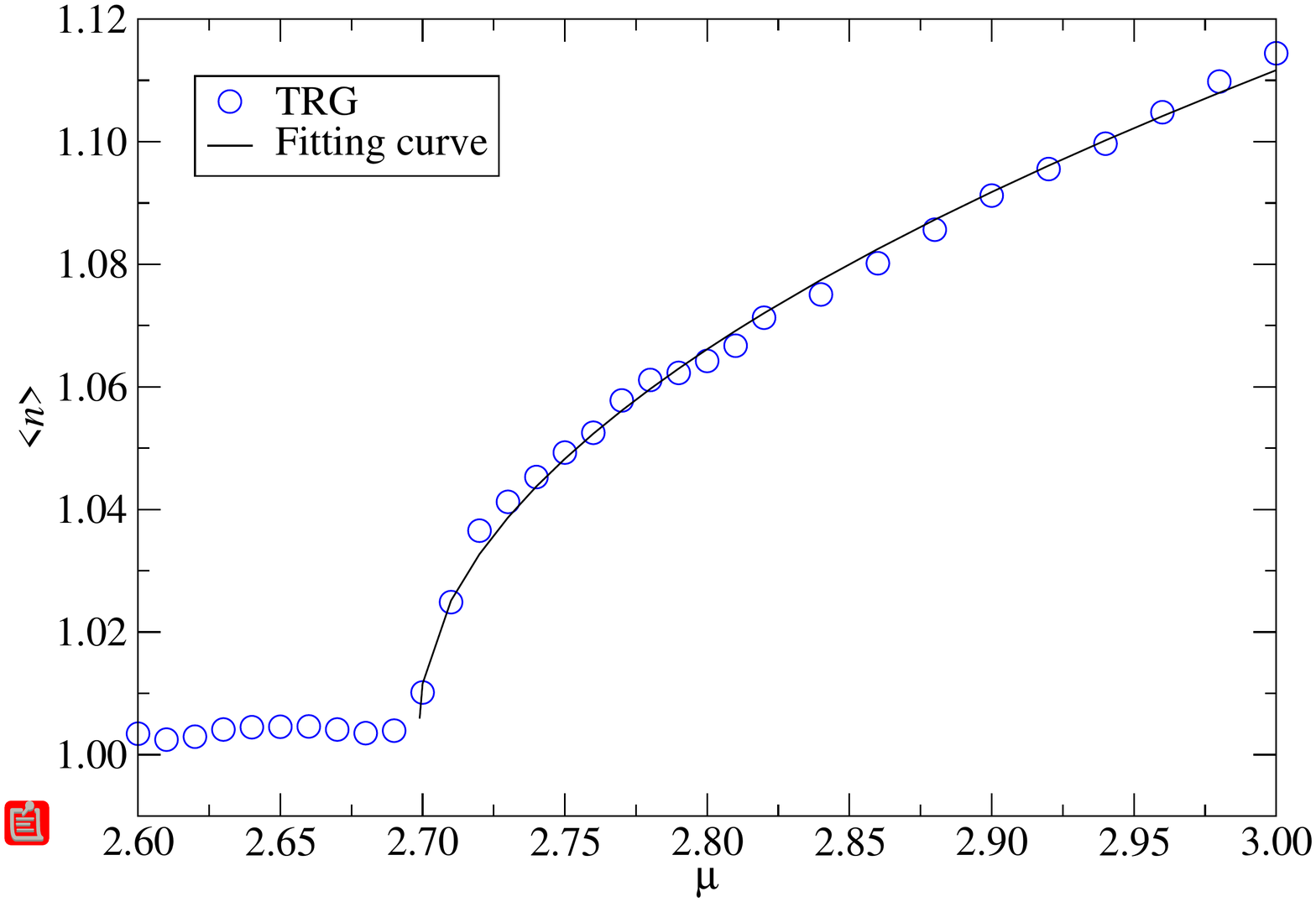}
		\caption{Electron density $\langle n\rangle$ at $\beta=1677.7216$ with $\epsilon=10^{-4}$ as a function of $\mu$. The bond dimension is chosen to be $D=80$.}
  		\label{fig:edensity}
  	\end{minipage}
  	\hspace*{3mm}
	\begin{minipage}[t]{0.48\hsize}
    		\centering
    		\includegraphics[width=1.0\hsize]{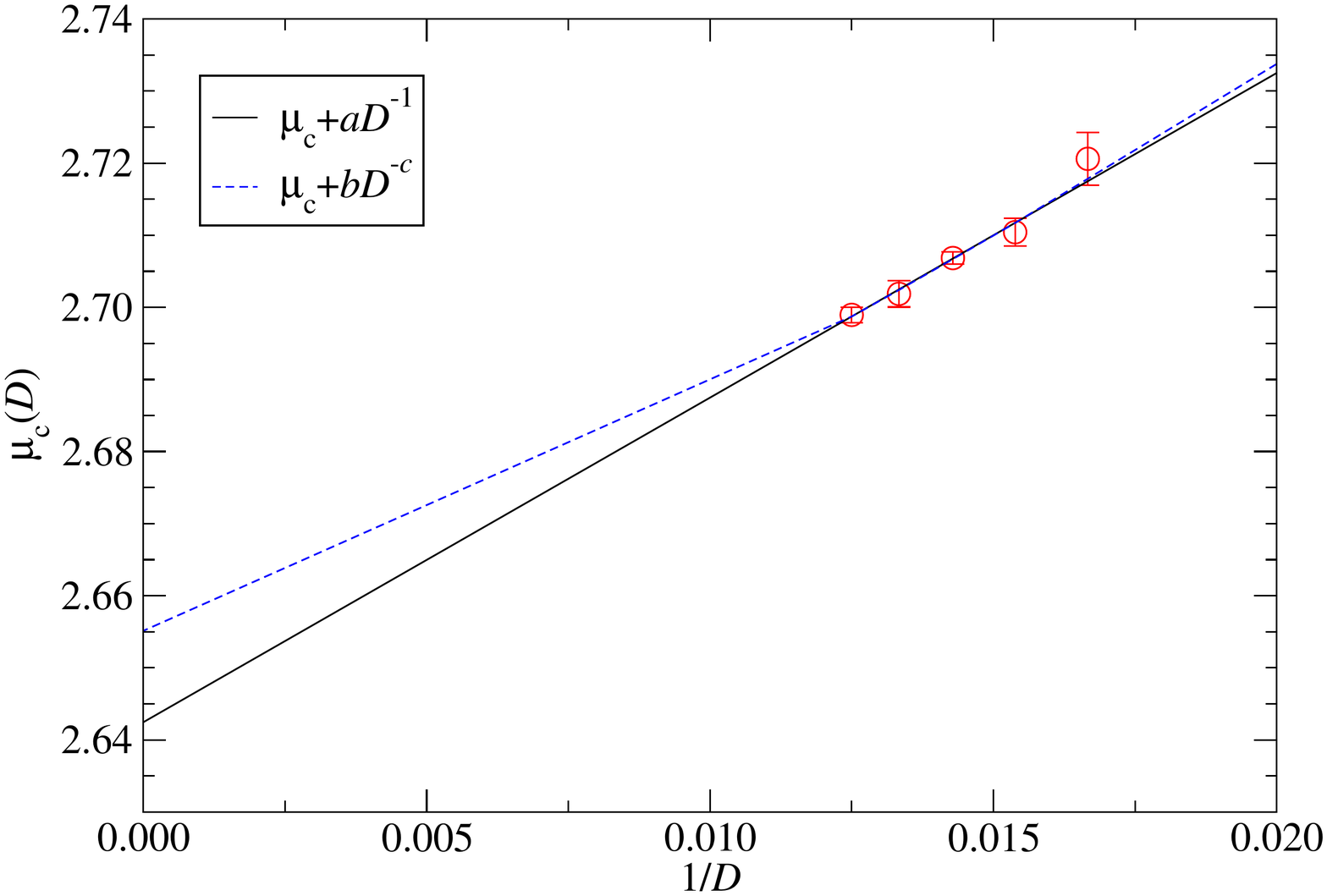}
		\caption{Critical chemical potential $\mu_{\rm c}(D)$ as a function of $1/D$. Solid line represents the fitting result with the function $\mu_{\rm c}(D)=\mu_{\rm c}+aD^{-1}$. Dotted curve also shows the fitting result with the function $\mu_{\rm c}(D)=\mu_{\rm c}+bD^{-c}$.}
 		\label{fig:mu_c}
	\end{minipage}
\end{figure}

We evaluate the electron density $\langle n\rangle$ following the numerical derivative in Eq.~(\ref{eq:nd_hubbard}).
Figure~\ref{fig:edensity} shows $\mu$ dependence of $\langle n\rangle$ near the criticality on $V=4096\times 1677.7216$ with $\epsilon=10^{-4}$ and $D=80$. The abrupt change of $\langle n\rangle$ at $\mu\approx 2.70$ indicates a metal-insulator transition. 
We determine the critical chemical potential $\mu_{\rm c}(D)$ and the critical exponent $\nu$ by fitting $\langle n\rangle$ in the metallic phase around the transition point with the following form:
\begin{align}
	\langle n\rangle=A+B\left|\mu-\mu_{\rm c}(D)\right|^\nu,
\end{align}
where $A$, $B$, $\mu_{\rm c}(D)$ and $\nu$ are the fit parameters.
The solid curve in Fig.~\ref{fig:edensity} shows the fitting result over the range of $2.68\le \mu\le 3.00$. We obtain $\mu_{\rm c}(D)=2.698(1)$ and $\nu= 0.51(2)$ at $D=80$.  Our result for the critical exponent is consistent with the theoretical prediction of $\nu=1/2$. 

\begin{table*}[htb]
	\caption{Critical chemical potential $\mu_{\rm c}(D)$ and critical exponent $\nu$ at each $D$.}
	\label{tab:mu_c}
	\begin{center}
		\begin{tabular}{|c|cccccc|}\hline
    		$D$ & 60 & 65 & 70 & 75 & 80 & $\infty$ \\ 
    		{\rm fit\; range} & [2.72,3.00] & [2.70,3.00] & [2.70,3.00] & [2.69,3.00] & [2.68,3.00] & $-$ \\ \hline
    		$\mu_{\rm c}(D)$ & 2.720(3) & 2.710(1) & 2.7068(8) & 2.701(1) & 2.698(1) & 2.642(05)(13)\\ 
    		$\nu$ & 0.49(3) & 0.52(1) & 0.50(2) & 0.51(2) & 0.51(2) & $-$\\ \hline
	\end{tabular}
	\end{center}
\end{table*}

In order to extrapolate the result of $\mu_{\rm c}(D)$ to the limit $D\to\infty$, we repeat the calculation changing $D$. The numerical results are summarized in Table~\ref{tab:mu_c}. In Fig.~\ref{fig:mu_c}, we plot $\mu_{\rm c}(D)$ as a function of $1/D$, providing two types of fittings. The solid line shows the fitting result with the function $\mu_{\rm c}(D)=\mu_{\rm c}+aD^{-1}$, which gives us $\mu_{\rm c}=2.642(5)$ and $a=4.5(4)$ with $\chi^2/{\rm d.o.f}=0.447093$. We have also fitted the data with the function $\mu_{\rm c}(D)=\mu_{\rm c}+bD^{-c}$, shown as the dotted curve in Fig.~\ref{fig:mu_c}, to estimate uncertainty in the choice of the fitting function. The difference between the central values of $\mu_{\rm c}$ obtained by these two types of fittings is considered to be a systematic error. Finally, we obtain $\mu_{\rm c}=2.642(05)(13)$ as the value of $\lim_{D\to\infty}\mu_{\rm c}(D)$, which shows good consistency with the exact solution of $\mu_{\rm c}=2.643\cdots$ based on the Bethe ansatz~\cite{PhysRevLett.20.1445,LIEB20031}. Our results show the efficiency of the TRG approach to the Hubbard model, being free from the sign problem. 

\subsection{(2+1)$d$ Hubbard model}
\label{subsec:hubbard_2+1}

Having succeeded in analyzing the (1+1)$d$ Hubbard model with the TRG method, we now investigate the doping-driven metal-insulator transition in the (2+1)$d$ case~\cite{Akiyama:2021glo}. Since its phase diagram is not well known so far, we calculate the electron density $\la n\ra$  as a function of the chemical potential $\mu$ choosing three values of the Coulomb potential with $U=80$, 8 and 2 as representative cases of the strong, intermediate and weak couplings. The $\mu$ dependence of $\la n\ra$ allows us to determine the critical chemical potential $\mu_{\rm c}$ at the doping-driven metal-insulator transition from the half-filling plateau with $\la n\ra=1$ to the metallic state with $\la n\ra>1$.

\subsubsection{Formulation and numerical algorithm}

The path-integral formulation for the partion function of the Hubbard model is already given in Sec.~\ref{subsubsec:hubbard_1+1_formulation}. The action in the $(2+1)d$ case is obtained by choosing $d=2$ in Eq.~(\ref{eq:d+1_action}). As in the $(1+1)d$ case, We employ the the periodic boundary condition in the spatial direction, $\psi(N_{x}+1,n_y,n_\tau)=\psi(1,n_y,n_\tau)$ and $\psi(n_x,N_{y}+1,n_\tau)=\psi(n_x,1,n_\tau)$, while the anti-periodic one in the temporal direction, $\psi(n_x,n_y,N_\tau+1)=-\psi(n_x,n_y,1)$.

The Grassmann tensor network representation of the partition function is obtained by following the procedure in Ref.~\cite{Akiyama:2020sfo}.
We evaluate the Grassmann tensor network generated by the rank-6 Grassmann tensor $\mathcal{T}_{\Psi_{x}\Psi_{y}\Psi_{\tau}\bar{\Psi}_{\tau}\bar{\Psi}_{y}\bar{\Psi}_{x}}$ employing the GATRG algorithm given in Ref.~\cite{Akiyama:2020soe}. As in the $(1+1)d$ case, after we carry out $m_{\tau}$ times of renormalization along with the temporal direction, the $3d$ ATRG procedure is applied as the spacetime coarse-graining. The optimal $m_{\tau}$ is found to be satisfying the condition $\epsilon 2^{m_{\tau}}\sim O(10^{-1})$ in the sense of preserved tensor norm.

\subsubsection{Numerical results}

\begin{figure}[htbp]
	\begin{minipage}[t]{0.48\hsize}
    		\centering
   		 \includegraphics[width=1.0\hsize]{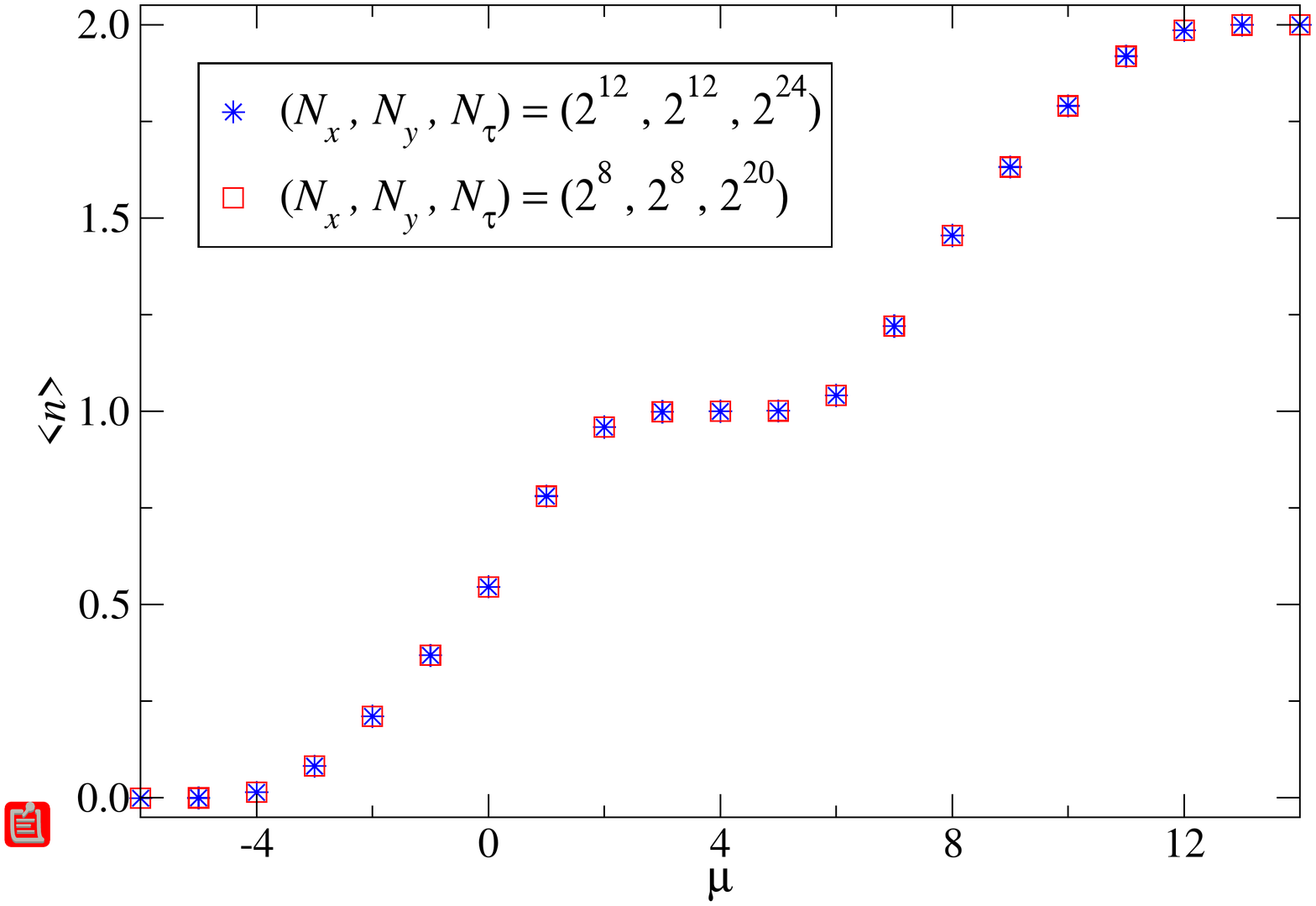}
   		 \caption{Electron density $\langle n\rangle$ at $U=8$ on two lattice sizes, $V=256^2\times 104.8576$ and $4096^2\times 1677.7216$, as a function of $\mu$. The bond dimension is set to be $D=80$.}
  		\label{fig:n_vol_2+1}
  	\end{minipage}
  	\hspace*{3mm}
	\begin{minipage}[t]{0.48\hsize}
   		 \centering
   		 \includegraphics[width=1.0\hsize]{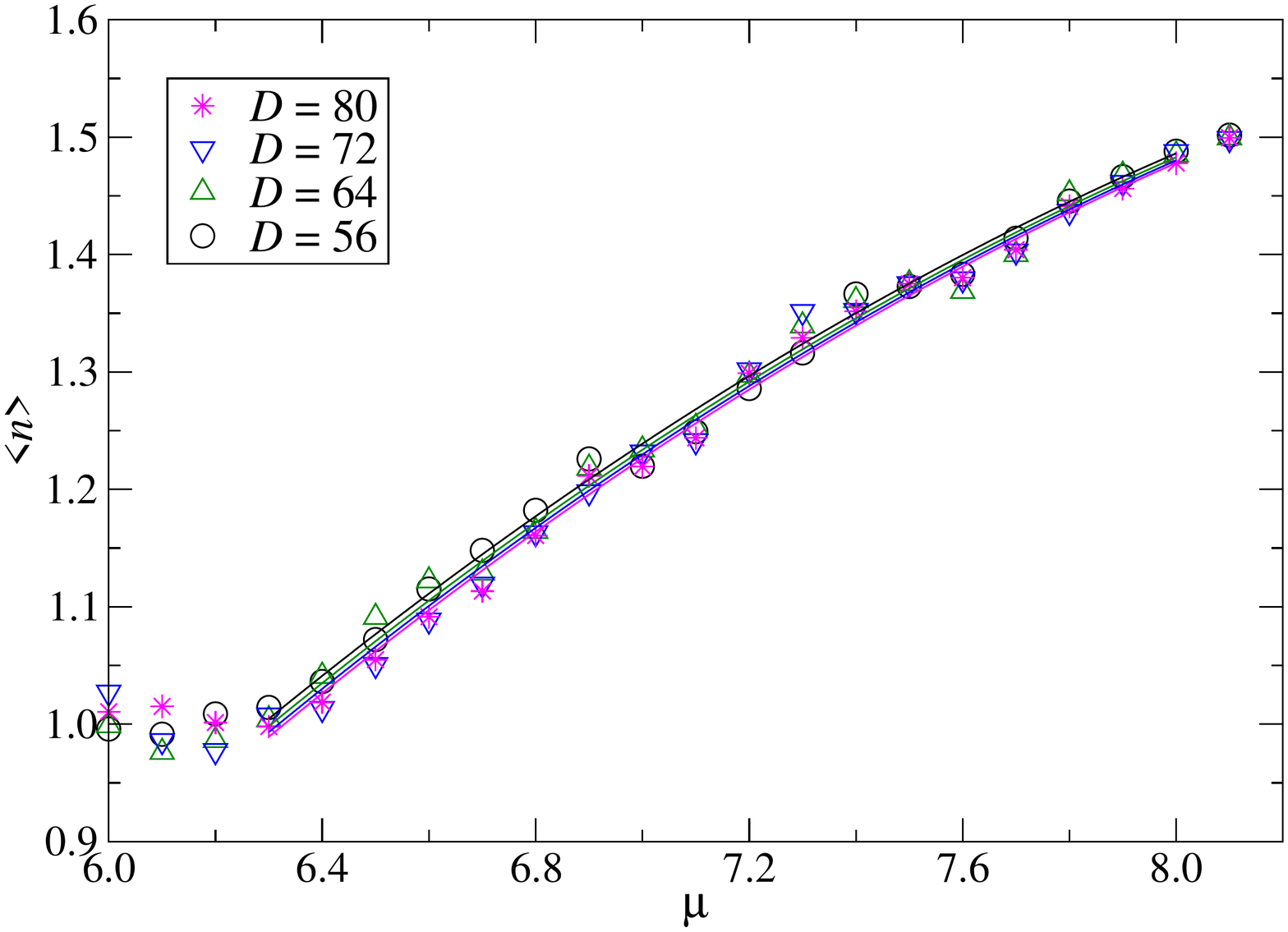}
   		 \caption{Electron density $\langle n\rangle$ at $U=8$ on $V=4096^2\times 1677.7216$  with $\epsilon=10^{-4}$ as a function of $\mu$. The bond dimensions are $D=80$, 72, 64 and 56. Fit results are drawn by the solid lines for each bond dimension.}
     		\label{fig:mu_c_2+1}
	\end{minipage}
\end{figure}

The $(U,t)=(8,1)$ case has been intensively investigated due to an expectation for a possible existence of the superconducting phase. In order to check the volume dependence of the electron density defined in Eq.~(\ref{eq:nd_hubbard}), we plot the $\mu$ dependence of $\la n\ra$ at $U=8$ in Fig.~\ref{fig:n_vol_2+1} changing the lattice sizes with $\epsilon=10^{-4}$, $m_{\tau}=12$ and $D=80$.  The results on $(N_{x},N_{y},N_{\tau})=(2^{8},2^{8},2^{20})$ and $(2^{12},2^{12},2^{24})$ are degenerate so that the latter lattice size, which corresponds to $V=4096^2\times 1677.7216$, is sufficiently large to be identified as the thermodynamic and zero-temperature limit. We observe the $\la n\ra=0$ plateau for $\mu\lesssim-4$ and the $\langle n\rangle=2$ one for $12\lesssim \mu$. The half-filling state is characterized by the plateau of $\langle n\rangle=1$ in the range of $2\lesssim \mu\lesssim 6$. These plateaus yield the vanishing compressibility $\kappa=\partial\langle n\rangle/\partial \mu$ indicating the insulating states.

Figure~~\ref{fig:mu_c_2+1} shows the $D$-dependence of  $\langle n\rangle$ around the metal-insulator transition with a much finer resolution of $\Delta \mu$ than Fig.~\ref{fig:n_vol_2+1} focusing on the range of $6.0\le \mu\le8.2$. The results at $D=80$, 72, 64 and 56 are almost degenerate indicating the small $D$ dependence. The critical chemical potential $\mu_{\rm c}$ is determined by the global fit with the following quadratic fitting function: 
\begin{align}
	\langle n\rangle=1+\alpha\left(\mu-\mu_{\rm c}(D)\right)+\beta\left(\mu-\mu_{\rm c}(D)\right)^2
\label{eq:n_fit}
\end{align}
with $\mu_{\rm c}(D)=\mu_{\rm c}(D=\infty)+\gamma/D$, where $\alpha$, $\beta$, $\gamma$ and $\mu_{\rm c}(D=\infty)$ are the fit parameters. The solid curves in Fig.~\ref{fig:mu_c_2+1} represent the fit results over the range of $6.3\le \mu\le8.0$. We obtain $\mu_{\rm c}(D=\infty)=6.43(4)$. 

We repeat the same analysis for the weak coupling case at $U=2$, whose critical chemical potential is found to be $\mu_{\rm c}(D=\infty)=1.30(6)$. The $\mu$ dependence of $\la n\ra$ in the strong coupling region is also investigated with the choice of $U=80$ at $D=80$. We obtain $\mu_c(D=80)=77.0(2)$ for the critical chemical potential. Our results at $U=80$, 8 and 2 show that $\vert \mu_{\rm c}- U/2\vert$ monotonically diminishes as $U$ decreases and seems to converge on $\vert \mu_{\rm c}- U/2\vert=0$ at $U=0$. This indicates the possibility that the model exhibits the metal-insulator transition at any finite $U$. This conclusion may provide us a different scenario of the phase diagram from that predicted by the dynamical mean-field theory (DMFT)~\cite{RevModPhys.68.13}; there exists some $U_{\rm c}$ such that no metal-insulator transition occurs with $U<U_{\rm c}$.

\section{Summary and outlook}
\label{sec:summary}

Since the application of the TRG method to QFTs was initiated in 2012, we have made a lot of progress in calculating the scalar, fermion, and gauge theories. We have developed efficient algorithms for various QFTs and have also shown that the TRG method is essentially free from the sign problem in the practical calculation. We are now able to investigate the 4$d$ scalar and fermionic theories. Aiming at the study of the finite density QCD, the only missing piece is an efficient algorithm to treat the non-Abelian gauge theories on higher ($\ge 3$) dimensions, whose development would be a primary task over the next few years.

\acknowledgments
 Numerical calculation for the present work was carried out with the supercomputer Fugaku provided by RIKEN (Project ID: hp200170, hp200314, hp210074, hp210204) and also with the Oakforest-PACS (OFP) and the Cygnus computers under the Interdisciplinary Computational Science Program of Center for Computational Sciences, University of Tsukuba.
This work is supported in part by Grants-in-Aid for Scientific Research from the Ministry of Education, Culture, Sports, Science and Technology (MEXT) (No. 20H00148) and JSPS KAKENHI Grant Number JP21J11226 (S.A.).

\bibliographystyle{JHEP}
\bibliography{bib/algorithm,bib/continuous,bib/discrete,bib/formulation,bib/gauge,bib/grassmann,bib/gravity,bib/others}

\end{document}